\documentclass[trackchanges]{aastex7}

\usepackage{amsmath}
\usepackage{booktabs}
\usepackage{multirow} 
\usepackage{hyperref}

\begin{document}

\title{Hemispheric Distribution of Solar Active Regions During Solar Cycles 23-25}

\author[orcid=0009-0009-5713-5380]{Yuxia Liu}
\affiliation{School of Mathematics and Computer Science, Yunnan Minzu University, Kunming, Yunnan, 650504, People’s Republic of China}
\email{18308802842@163.com}  

\author[orcid=0000-0002-9997-9524]{Tingting Xu} 
\affiliation{School of Mathematics and Computer Science, Yunnan Minzu University, Kunming, Yunnan, 650504, People’s Republic of China}
\email{xutingting@ymu.edu.cn}

\author[0000-0001-8865-1497]{Miao Wan}
\affiliation{School of Mathematics and Computer Science, Yunnan Minzu University, Kunming, Yunnan, 650504, People’s Republic of China}
\email{wanmiao@ymu.edu.cn }

\author[0000-0003-4407-8320]{Linhua Deng}
\affiliation{School of Mathematics and Computer Science, Yunnan Minzu University, Kunming, Yunnan, 650504, People’s Republic of China}
\email{linhua.deng@ymu.edu.cn}

\author[0000-0002-9977-2646]{Xinhua Zhao}
\affiliation{State Key Laboratory of Space Weather, National Space Science Center, Chinese Academy of Sciences, Beijing 100190, People’s Republic of China}
\affiliation{Radio Science and Technology Center  $(\pi~\text{Center})$, Chengdu, Sichuan, 610041, People’s Republic of China}
\email{xhzhao@spaceweather.ac.cn}

\author[0000-0002-6957-8009]{Shiyang Qi}
\affiliation{College of Science, Henan Agricultural University, Zhengzhou, Henan, 450002, People’s Republic of China}
\email{qishiyang@henau.edu.cn}

\author[0000-0001-9062-7453]{Nanbin Xiang}
\affiliation{Yunnan Observatories,  Chinese Academy of Sciences, Kunming, Yunnan, 650216, People’s Republic of China}
\email{nanbin@ynao.ac.cn}

\author[0009-0003-2287-6441]{Weihong Zhou}
\affiliation{School of Mathematics and Computer Science, Yunnan Minzu University, Kunming, Yunnan, 650504, People’s Republic of China}
\affiliation{Key Laboratory for the Structure and Evolution of Celestial Objects, Chinese Academy of Sciences, Kunming, Yunnan, 650216, People’s Republic of China}
\email{zwh@ymu.edu.cn}

\correspondingauthor{Weihong Zhou}
\email{zwh@ymu.edu.cn}

\begin{abstract}
Solar active regions (ARs) are crucial for understanding the long-term evolution of solar activities and predicting eruptive phenomena, including solar flares and coronal mass ejections. However, the cycle-dependent properties in the north-south asymmetry of ARs have not been fully understood. In this study, we investigate the hemispheric distribution of ARs from Carrington Rotation 1909 to 2278 (between 1996 May and 2023 November) by using three parameters that describe the magnetic field distribution of ARs: number, area, and flux. The main findings are as follows: (1) The three AR parameters show significant hemispheric asymmetry in cycles 23-25. The strong correlation between AR area and flux indicates that they can better reflect the intrinsic properties of solar magnetic field. (2) The correlation between sunspot activity and AR parameters varies in the two hemispheres across the different cycles. The AR parameters provide additional information for the variations in sunspot activity, which can better predict the intensity and cyclical changes of solar activity. (3) The variation in the fitting slope sign of the asymmetry index for AR parameters reflects periodic changes in hemispheric ARs, providing valuable insights into the activity of other stars. (4) Both the dominant hemisphere and the cumulative trend of AR parameters display a cycle-dependent behavior. Moreover, the trend variations of AR area and flux are similar, reflecting the long-term evolutionary characteristics of solar magnetic field. Our analysis results are relevant for understanding the hemispheric coupling of solar magnetic activity and its cyclic evolutionary patterns.

\end{abstract}


\section{Introduction} 
Active regions (ARs) are areas on the Sun characterized by strong magnetic fields, originating from the emergence of flux tubes \citep{2009LRSP....6....4F,2010ApJ...720..233C}. They are regarded as the primary sources of various solar activities ranging from small-scale brightenings and jets to major events such as flares and coronal mass ejections (CMEs) \citep{2015LRSP...12....1V}. These regions determine solar polar fields and drive the variability of the solar cycle, exerting different impacts on space weather and various solar activities \citep{2017SoPh..292..167N,2019ApJ...871...16J,2019LRSP...16....3T}. Therefore, studying the temporal and spatial behaviors of ARs is particularly important. Lots of studies have investigated the formation and evolution \citep{1981sars.work...17S,2010ApJ...720..233C,2012ApJ...757..147C,2012ApJ...753L..13S,2023SSRv..219...63W}, magnetic properties \citep{2016ApJ...820L..11J,2018Ge&Ae..58.1159A} of ARs, revealing flux variations and inter-cycle differences \citep{2019A&AT...31...75Z,2020Ge&Ae..60..673Z,2022Ge&Ae..62..823Z,2023SSRv..219...64N}. These studies provide both theoretical and observational support for understanding the mechanisms and evolution of solar magnetic activity \citep{2014ARA&A..52..251C,2021LRSP...18....2C}. 

The study of individual AR and their properties began with sunspot drawings and daily observations, but was revolutionized by \citet{1908ApJ....28..315H}, who demonstrated their magnetic nature (Hale's law). The interactions of rising flux concentrations and supergranule-scale flows during the emergence process are key physical mechanisms in AR formation \citep{1965ApJ...141.1492B,2019A&A...628A..37B}. The short-lived ARs exhibit regularity at different latitudes and evolve with solar cycles \citep{1988SvA....32..211K}. The magnetic properties of AR display significant north-south (N-S) asymmetry during the cycle maximum, which is related to the internal magnetic field generation mechanisms of the Sun and magnetohydrodynamic (MHD) processes \citep{1998ASPC..140..131C}. In recent years, significant progress has been made in AR research. \citet{2022STP.....8d..28G} found that the formation of ARs is closely related to the large-scale magnetic field. \citet{2024ApJS..272....5Z} analyzed automatically detected ARs and found that the northern hemisphere dominated both in AR number and cumulative area during solar cycle 24 (SC24). By studying ARs of different morphologies, \citet{2019A&AT...31...75Z,2020Ge&Ae..60..673Z} revealed the interaction between mean-field dynamo mechanisms and small-scale turbulent dynamo processes, as well as the magnetic properties and temporal evolution of ARs across solar cycles. Subsequently, \citet{2023AdSpR..71.1984Z,2024MNRAS.532.2032Z} found that ARs of different morphologies exhibit a significant N-S asymmetry that varies over the solar cycles. The hemispheric asymmetry may be attributed to dynamo processes of the global magnetic field, the interaction of dipolar and quadrupolar components, and turbulent magnetic field effects. Although numerous studies have explored the characteristics of ARs, there is still a lack of extensive research on their asymmetric distribution. To date, the study of the N-S asymmetry in solar activity has garnered considerable attention \citep{2001SoPh..199..211D,2007MNRAS.381.1527J}. 

The N-S asymmetry is one of the most striking manifestations of the solar cycle \citep{2018A&A...618A..89S,2022MNRAS.511..472D}, driven by the dynamic processes of the convective dynamo \citep{2019MNRAS.489.4329H,2021ApJ...919...36K,2021SoPh..296...86B}. The hemispheric asymmetry of the solar magnetic structures plays a crucial role in understanding the internal dynamo mechanism and sub-photospheric dynamics \citep{2013ApJ...768..188C}. The 11-year cycle of sunspot positions follows the ``Spörer’s Law of Zones'' as illustrated by the well-known ``Butterfly Diagram'' \citep{1904MNRAS..64..747M}. This periodicity is reflected not only in the variation of sunspot numbers \citep{1999JGR...10422375H}, but also in the spatial distribution of ARs \citep{2020Ge&Ae..60..673Z}.

Many previous studies have used solar activity indicators, including sunspot numbers and sunspot group numbers \citep{2000MNRAS.317..897L,2002A&A...390..707T,2009SoPh..254..145L,2021A&A...652A..56V}, sunspot area \citep{1994SoPh..152..481O,2005A&A...431L...5B,2021SoPh..296....2R}, flare index \citep{1996SoPh..166..201A,2001SoPh..198..399A,2003SoPh..214..375O,2004SoPh..223..287O,2020SoPh..295..100R}, filaments/prominences \citep{2000SoPh..194...87V,2010NewA...15..346L,2024ApJS..272....5Z}, and coronal mass ejections (CMEs) \citep{2009MNRAS.400.1383G,2023MNRAS.520.3923Z} to study the asymmetry of solar activity. For solar ARs, most studies have focused primarily on magnetic flux to analyze N-S asymmetry. However, other parameters, such as the number and area of AR, are also critical indicators of solar magnetic activity \citep{1984SoPh...91...75T,2015LRSP...12....4H}. The differences and evolutionary patterns of these parameters remain inadequately understood between the two hemispheres and different solar cycles. Thus, more research is needed to better understand the hemispheric characteristics of ARs in different parameters. 

Moreover, past studies and reconstructions of solar activity have utilized data from all magnetic regions on the solar surface (i.e. on dark sunspots, bright faculae-plage and network) \citep{2020SoPh..295...38B,2021ApJ...920..100W,2022A&A...667A.167C,2024ApJ...976...11P}. These data are crucial for understanding the magnetic activity and variations of the Sun and Sun-like stars. The sunspot number is the only direct measurement of solar activity (approximately 1600-1900) \citep{2003PhRvL..91u1101U,2020LRSP...17....1A}, and many studies have used it as the sole proxy to reconstruct past magnetic activity and irradiance \citep{2021GMS...258...83P,2023SSRv..219...31Y}. These studies improved our understanding of the mechanisms behind the solar activity variations. Therefore, studying the relationship between AR parameters (i.e. number, area and flux) and sunspot activity from 1996 to 2023 is also important for reconstructing past solar activity.

In this paper, we use the latest comprehensive AR database provided by \citet{2024ApJ...971..110W} that systematically analyzes the hemispheric distribution characteristics of AR parameters during SC23 to SC25. In addition, we further study the relationship between the AR parameters and sunspot activity, providing a new perspective for solar magnetic activity. The remainder of this paper is structured as follows: Section \ref{sec:Data} introduces the data and methods used in our study. Section \ref{sec:AnR} presents the results and analysis of the hemispheric distribution characteristics of ARs. Conclusions and discussions are provided in Section \ref{sec:CD}.

\section{Data and Methods}
\label{sec:Data}
\subsection{AR Parameters}
\label{ar_preprocessing}
The data for this study are derived from the live homogeneous AR database\footnote{\url{https://github.com/Wang-Ruihui/A-live-homogeneous-database-of-solar-active-regions/tree/with-repeat-AR-Removal-and-params-}
\url{dipole-fields}} provided by \citet{2023ApJS..268...55W,2024ApJ...971..110W}. They developed a method to automatically detect ARs from 1996 onward based on the Michelson Doppler Imager on board the Solar and Heliospheric Observatory (SOHO/MDI) \citep{1995SoPh..162..129S} and the Helioseismic and Magnetic Imager on board the Solar Dynamics Observatory (SDO/HMI) \citep{2012SoPh..275..207S} synoptic magnetograms. This method has advantages in excluding decayed ARs, unipolar regions and ensuring compatibility with any available synoptic magnetograms. They calibrated the identified AR area and flux and calculated the dipole field. This database enhances our understanding of solar cycles, particularly the variations in the polar magnetic field.

The database comprises 2892 ARs spanning from CR 1909 to CR 2278 (1996.5–2023.11), includes two sets of parameters: the first set consists of basic AR parameters, such as number, area and total unsigned flux, while the second set includes parameters directly related to solar cycle variability, such as the initial dipole field ($D_i$), the final dipole field ($D_f$), and the bipolar magnetic region (BMR) approximations for both parameters, denoted as $D_i^B \text{ and } D_f^B$. This study employs the following key parameters: the latitude of the flux-weighted centroid for both polarities and the whole AR, and the area and flux of each polarity. Additionally, the Carrington Rotation (CR) number and label are commonly used to uniquely identify each AR. We apply the CR period as defined by \citet{2006A&A...449..791T} to calculate the specific appearance times of each AR.

\begin{figure*}[ht!]
\centering
\begin{minipage}[t]{0.51\textwidth} 
  \centering
  \includegraphics[width=\textwidth, height=12cm]{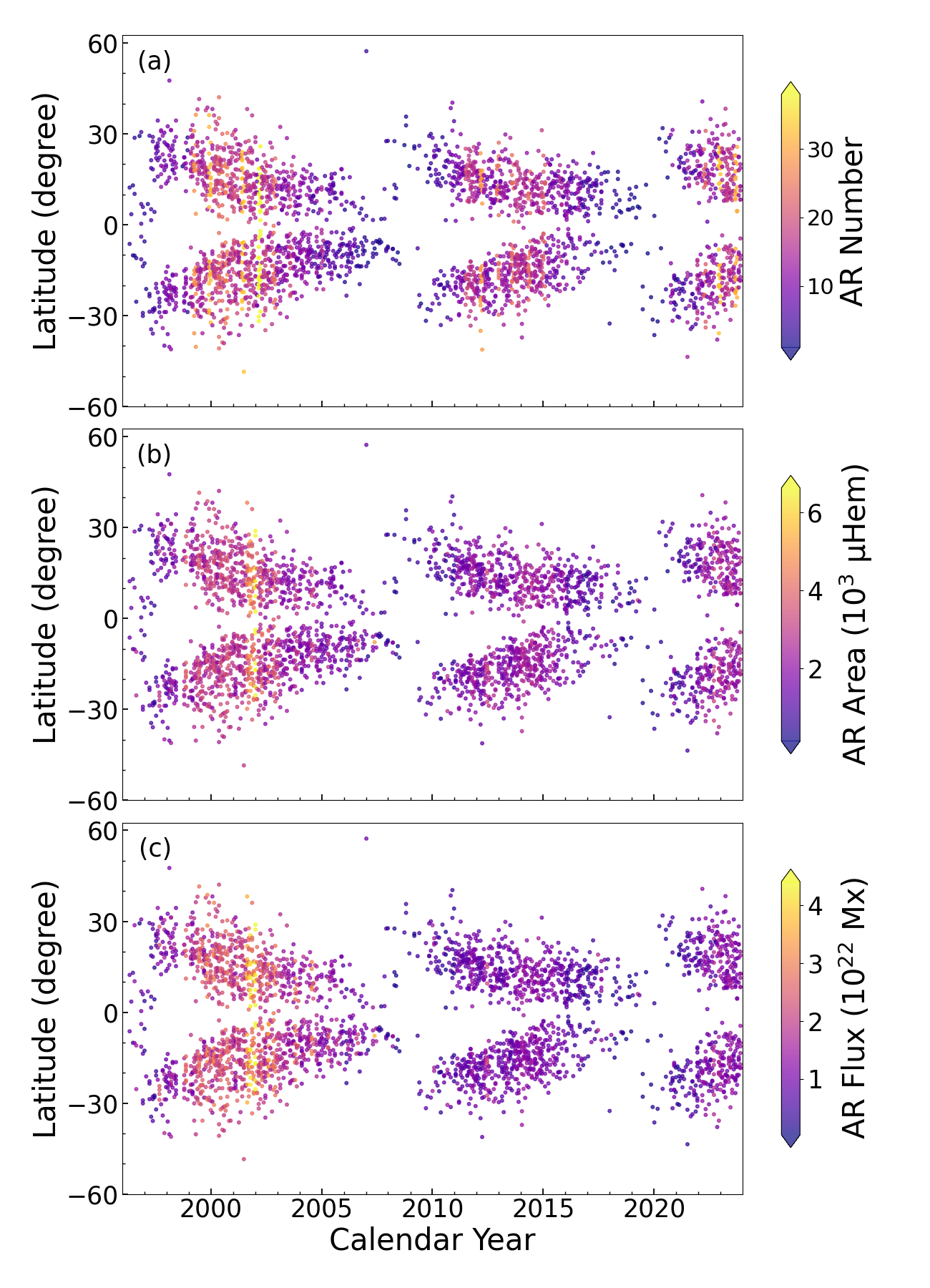}
  \caption{Butterfly diagrams of AR parameters from 1996 May to 2023 November. From top to bottom, the panels depict the latitude-time distribution of AR number, area, and flux. The colors show AR number, average area, and flux, respectively.\label{fig:lat}}
\end{minipage}
\hfill
\begin{minipage}[t]{0.46\textwidth}
\centering
\includegraphics[width=\textwidth, height=12cm]{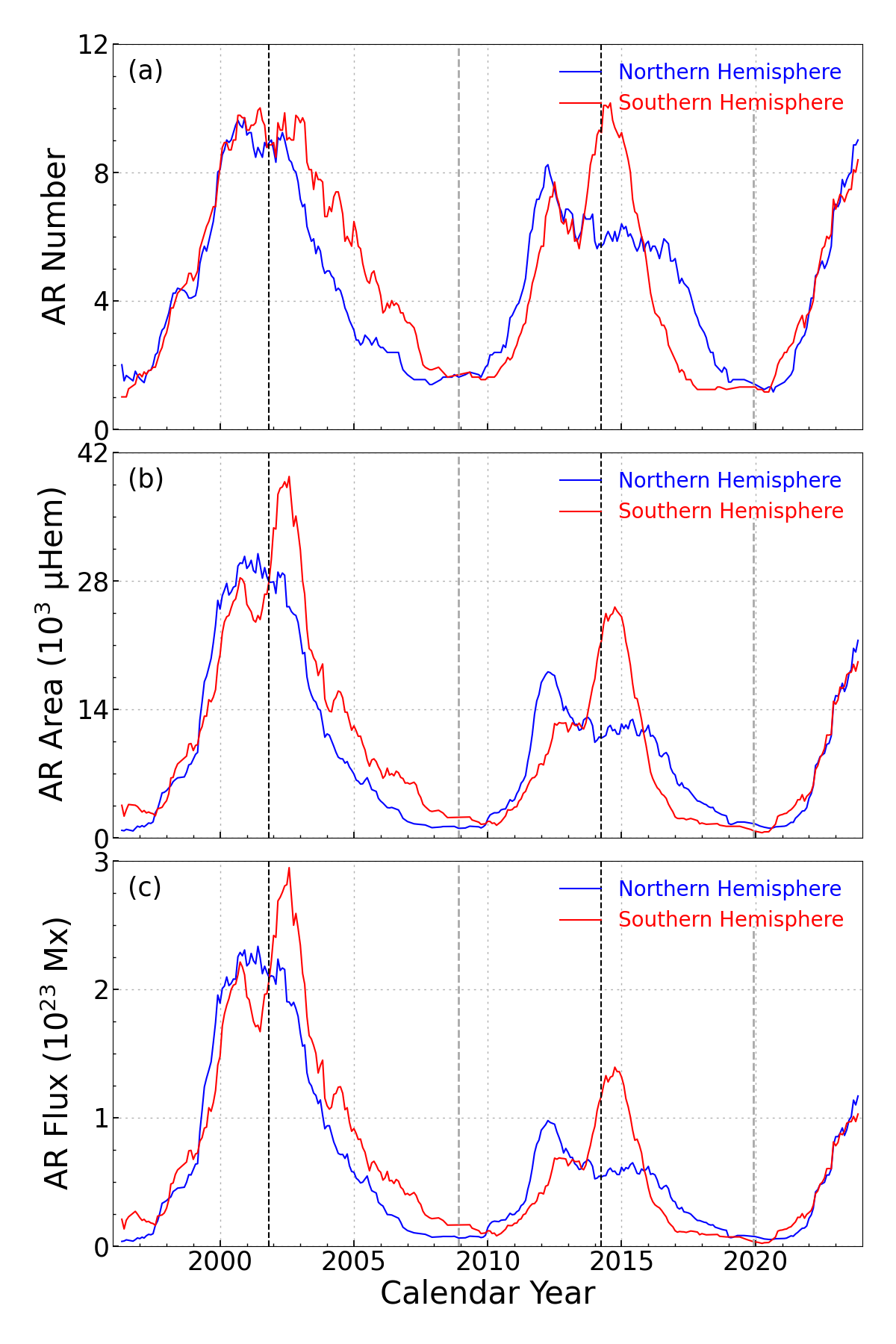}
\caption{Time evolution of the AR parameters from 1996 May to 2023 November. From top to bottom: AR number, area, and flux. The blue (red) lines represent the northern (southern) hemisphere. The black (gray) dashed lines represent the maximum and minimum times of the solar cycles.\label{fig:ns}}
\end{minipage}
\end{figure*}

As illustrated in Figure \ref{fig:lat}, the latitude-time relationship diagrams of the three AR parameters from 1996 to 2023 form the typical pattern ``Maunder butterfly diagram'' \citep{1904MNRAS..64..747M}. Panels (a), (b), and (c) clearly demonstrate that: (1) ARs begin emerging at mid-latitudes (approximately 30 degrees) in the early stages of the cycle. As the cycle progresses, the emergence moves closer to the equator. This equatorward progression of ARs emergence is known as Spörer’s Law \citep{1903Obs....26..329M}; and (2) the spatial and temporal variations of ARs are asynchronized between the two hemispheres. 

Figure \ref{fig:ns} displays the temporal variations of the three AR parameters in different hemispheres, spanning SC23 to part of SC25. All data are processed with a 13-month smoothing filter to remove high-frequency fluctuations. The curve reveals periodic fluctuations in solar activity and highlights differences between the two hemispheres. From Figure \ref{fig:ns}, one can see that: (1) there is a clear N-S asymmetry in the three parameters of AR across different solar cycles; (2) the trends of AR area and flux are similar throughout the period that show significant differences from the AR number; and (3) the AR area and flux in SC24 have decreased significantly compared to SC23, while the AR number remains relatively consistent across both cycles. These differences may offer valuable insights for predicting the intensity of future solar cycles.

\subsection{Sunspot Parameters}
\label{fas}
ARs are concentrated areas where the solar magnetic field interacts with plasma and serve as crucial environmental factors in the formation of sunspots. Sunspot activity is a fundamental indicator of the solar cycle \citep{1844AN.....21..233S,2015LRSP...12....4H}, and the numbers and areas of sunspot are critical observational metrics for assessing magnetic field strength. Studying the relationship between sunspot activity and AR parameters may provide important values for understanding solar internal magnetic activity and dynamic processes.

\begin{figure*}[ht!]
\centering
\includegraphics[width=0.8\textwidth]{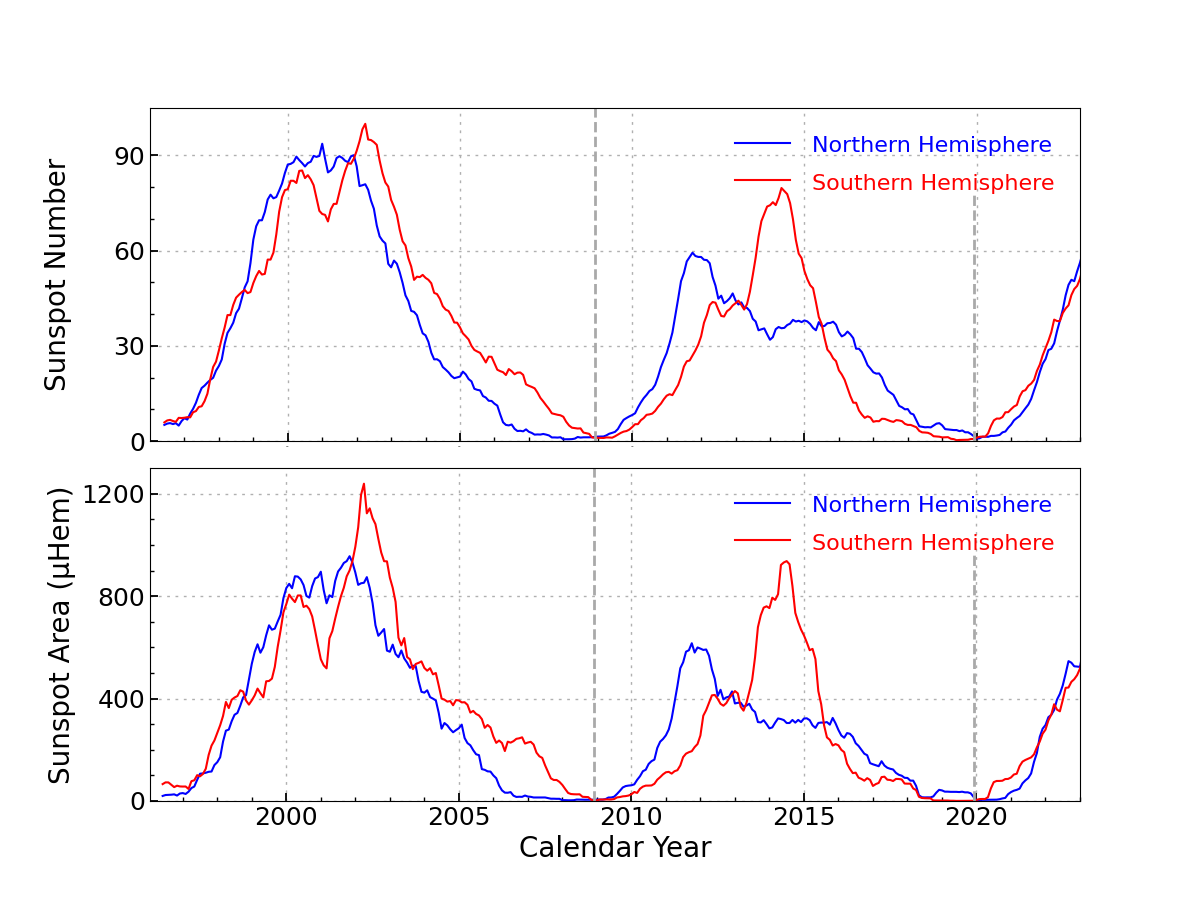}
\caption{Monthly distribution of sunspot numbers and areas. Top panel: the monthly average hemispheric sunspot numbers, with the northern (southern) hemisphere show in blue (red); Bottom panel: same as the top panel, but corresponds to sunspot areas, smoothed using a 13-month running average. The gray dashed lines represent the minimum times of the solar cycle.\label{fig:ssn-sa}}
\end{figure*}

This study utilizes the monthly time series of the hemispheric sunspot data, compiled by the World Data Center - Sunspot Index and Long-term Solar Observations (WDC-SILSO)\footnote{\url{https://www.sidc.be/SILSO/datafiles}} 
\citep{2014SSRv..186...35C}. The database has been widely regarded as the standard reference for solar cycle behavior and is commonly used to investigate the long-term evolution of the solar cycle \citep{2020LRSP...17....1A,2021MNRAS.504.5199C}. The sunspot area data are derived from the monthly sunspot areas (in units of millionths of a solar hemisphere) \citep{2015LRSP...12....4H} compiled by the Royal Greenwich Observatory (RGO), the US Air Force (USAF), and the National Oceanic and Atmospheric Administration (NOAA)\footnote{\url{http://solarcyclescience.com/activeregions.html}}. Figure \ref{fig:ssn-sa} presents the monthly distribution of sunspot numbers and areas across different hemispheres from 1996 May to 2023 November, aligned with the specific occurrences of ARs.

As shown in Figure \ref{fig:ssn-sa}, there are significant differences in solar activity between the two hemispheres, as can be observed: (1) the sunspot numbers and areas occur asynchronously in both hemispheres, exhibiting different intensities of development; and (2) the sunspot numbers and areas in SC24 have decreased significantly compared to SC23, especially in the northern hemisphere. This trend resembles the changes for the AR area and flux shown in Figure \ref{fig:ns}, but differs from the variation of the AR number.

\subsection{Analysis Methods} 
\label{hemispheric asymmetry}
\subsubsection{Expression of Hemispheric Asymmetry}
\label{expression_asymmetry}
The hemispheric asymmetry of solar activity is one of the most significant and intriguing features of the solar cycle. The N-S asymmetry of AR parameters can be characterized by the normalized asymmetry index ($NAI$) \citep{1955MNRAS.115..398N,2005A&A...431L...5B} or the absolute asymmetry index ($AAI$) \citep{2006A&A...447..735T,2006A&A...449L...1Z,2007A&A...475L..33D}, which are expressed as follows:

\begin{equation}
\label{eq:1}
\textit{NAI} = \frac{\textit{V}_{N} - \textit{V}_{S}}{\textit{V}_{N} + \textit{V}_{S}},
\end{equation}

\begin{equation}
\label{eq:2}
\textit{AAI} = \textit{V}_{N} - \textit{V}_{S},
\end{equation}

\noindent Here, $\textit{V}_{N}$ and $\textit{V}_{S}$ represent the values of three AR parameters in a certain year for the northern and southern hemispheres, respectively.
 
Furthermore, to better illustrate the long-term behavior of the hemispheric asymmetry, \citet{2009SoPh..254..145L} proposed the cumulative absolute asymmetry index ($CAAI$) \citep{2015AJ....150...74Z,2024ApJ...962..172Z}, defined as follows:

\begin{equation}
\label{eq:3}
\textit{CAAI}= \ \text{\textit{C}}_N - \text{\textit{C}}_S.
\end{equation}

\noindent Here, $\text{\textit{C}}_N$ and $\text{\textit{C}}_S$ denote the cumulative monthly values of AR parameters in the northern and southern hemispheres, respectively. We calculated the values of AR parameters for the two hemispheres during SC23 to SC25, as shown in Tables \ref{tab:numbers} to \ref{tab:all}. By quantitatively analyzing these asymmetry indices, we can gain deeper insights into the magnetic field evolution dynamics between the
two hemispheres across different cycles and reveal their underlying physical mechanisms.

\subsubsection{Statistical Testing}
To assess the statistical significance of the normalized asymmetry index for each year, we calculated the actual probability of generating such an asymmetrical distribution. Consider \(n\) objects distributed across two hemispheres, where each object has a random probability of $p = \frac{1}{2}$ appearing in one hemisphere. We use the following binomial formula~\eqref{eq:4} to derive the probability \(P(k)\) of obtaining \(k\) objects in one hemisphere and \((n - k)\) objects in the other hemisphere \citep{1990A&A...229..540V,1993A&A...274..497C,2006JApA...27..151J}.

\begin{equation}
\label{eq:4}
P(k) = \binom{n}{k} p^k (1 - p)^{n - k} 
\end{equation}

\noindent The probability of obtaining more than \(d\) objects in one hemisphere is:
\begin{equation}
\label{eq:5}
P(\geq d) = \sum_{k=d}^{n} P(k) 
\end{equation}

In general, if the probability of AR parameters in a certain year is less than 10\% or greater than 90\%, it is deemed that the distribution of ARs in the two hemispheres has statistical significance and is not caused by random fluctuations. In contrast, if the probability is greater than 10\% and less than 90\%, the distribution of ARs in both hemispheres is relatively balanced and statistically insignificant.

\section{Analysis Results}
\label{sec:AnR}
\subsection{Distribution of AR Parameters} \label{subsec:b}
As illustrated in Figure \ref{fig:ns}, the three AR parameters show different temporal behaviors in the two hemispheres. For AR number, the two hemispheres are highly synchronized, and the distributions remain relatively concentrated during the ascending phase of SC23. However, significant phase differentiation occurs between the two hemispheres during the declining phase of SC23, with the southern hemisphere leading the northern one. In addition, the activity patterns of both hemispheres exhibit a step-like evolution. The depth of the gap between the two main maxima of the cycle in different hemispheres is similar, and the time profiles look almost like a plateau. After entering SC24, the two main maxima are formed by ARs in different hemispheres (the northern hemisphere dominates in the first maximum, while the southern hemisphere takes the lead in the second maximum), reflecting a marked hemispheric difference. This trend is observed not only in the AR number but also in the variations of AR area and flux. This aligns with the findings of \citet{2016ApJ...830L..33M}, who showed that the double peaks may occur in one of the hemispheres without having any counterpart of the same in the other hemisphere. During the ascending phase (2010-2013) and the descending phase (2016-2020) of SC24, the activity of the northern hemisphere significantly exceeds that of the southern one. Furthermore, it should be noted that activity in the southern hemisphere increases rapidly and forms the second peak after the first peak of SC24, while the northern hemisphere weakens and levels off. This divergence results in a significant amplitude difference between the two hemispheres. 

For AR area and flux, their variation trends are quite similar but differ from the behavior of AR number. The distribution of activity across the two hemispheres is relatively dispersed, with a slight overall dominance in the northern hemisphere during the ascending phase of SC23. The first peak occurs in the northern hemisphere, followed by a rapid increase in AR area and flux in the southern hemisphere. Meanwhile, the northern hemisphere's activity enters a relatively steady phase, displaying a plateau-like structure. During this interval, the AR number in both hemispheres remains relatively stable. This phenomenon can be attributed to the turbulent component of the dynamo, which leads to variations in the magnetic flux of AR, thereby enhancing activity in the southern hemisphere. This suggests that variations in AR area and flux are more critical for studying the solar surface magnetic field.
 
The activity in the southern hemisphere is slightly stronger than in the northern one during SC25 (2021-2022). Subsequently, the activity becomes relatively concentrated across both hemispheres, and the northern hemisphere leads to the southern one by 2023 November. This phase of evolution provides valuable insights for understanding and predicting solar cycles.

Overall, the AR parameters show significant asynchrony during different solar cycles. In addition, compared to SC23, there is a significant decrease in AR area and flux during SC24, which may be attributed to the intensity of the solar cycles \citep{2020ApJ...903..118D,2022AdSpR..69..798P}. The trends in AR area and flux are more pronounced than the AR number, a fact that had been studied before \citep{2023AdSpR..71.1984Z}. The double-peak structure of a cycle as a whole is formed by ARs from both hemispheres, except for the AR number in SC23. This asynchrony reveals an asymmetric evolution of the solar magnetic field between the two hemispheres, reflecting the complex magnetic field restructuring and hemispheric differences throughout the solar cycle. 

\subsection{Correlations among AR Parameters}
\label{AR parameters}
According to the study by \citet{2023ApJS..268...55W}, calibration of parameters based on SOHO/MDI and SDO/HMI synoptic magnetograms found that discrepancies still persisted after the corrected identification results. This mainly affected the AR number with minimal impact on the AR area and flux. They emphasized that AR area and flux are critical for studying the solar surface magnetic field, whereas the AR number is relatively less impactful. Furthermore, they also found that the ratios of SC24 to SC23 for AR area and flux are almost identical. This similarity is consistent with the well-known linear correlation between the AR area and flux \citep{1966ApJ...144..723S}.

As mentioned in Section~\ref{subsec:b}, the trend of variation of the AR area and flux are similar between the two hemispheres. Therefore, the correlations among the three AR parameters may provide deeper insights, and it is more likely to help further understand the roles and mutual interactions of different parameters during solar cycles, as well as the formation and evolution of the solar magnetic field. 

\begin{figure*}[ht!]
\centering
\includegraphics[width=0.9\textwidth]{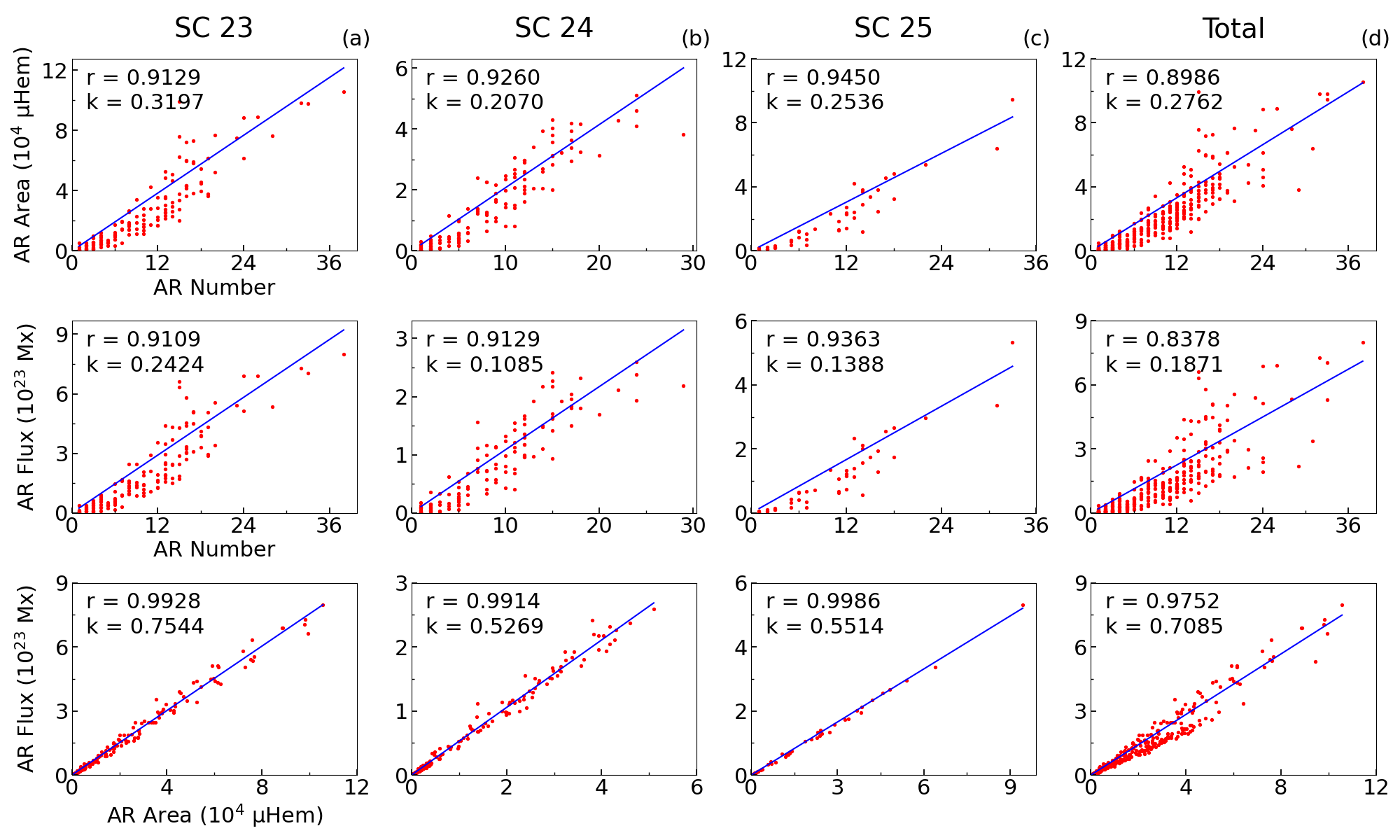}
\caption{Scatter plots of correlation coefficients between pairs of AR parameters across the different solar cycles. From top to bottom, the panels display the correlation between AR number and area, AR number and flux, and AR area and flux, with each subplot labeled with the respective correlation coefficient (\protect\( r \protect\)) and the slope of the fitted line (k). Red dots indicate actual observed data points, while the blue line represents the fitted trend line.}
\label{fig:cor}
\end{figure*}

Figure \ref{fig:cor} shows the correlation between the three AR parameters across the different solar cycles. It is evident from the figure that AR area and flux consistently exhibit the strongest correlation across all observed periods, with the Pearson correlation coefficients reaching 0.9928, 0.9914, 0.9986, and 0.9752 for SC23, SC24, SC25 and the entire period, respectively. The correlation coefficient in Figure \ref{fig:cor} are both above the confidence level 95\%, implying a high degree of correlation between the three AR parameters. This robust correlation is succeeded by the relationship between AR number and area, while the correlation between AR number and flux remains the lowest. This relationship remains consistent across the different solar cycles, further validating and reinforcing the conclusions of \citet{1966ApJ...144..723S}, providing a better understanding of the solar magnetic field's variation patterns and its behavior across the different solar cycles. In addition, we also observe that compared to SC23, the correlation between AR area and flux decreases during SC24, while the correlation between AR number and area as well as between AR number and flux increased, highlighting the differences among the three parameters of AR.

\subsection{Differences between AR Parameters and Sunspot Parameters}
\label{AR parameters and solar activity}
\begin{figure}[htbp]
\centering
\begin{minipage}[t]{0.49\textwidth} 
  \centering
  \includegraphics[width=\textwidth, height=12.5cm]{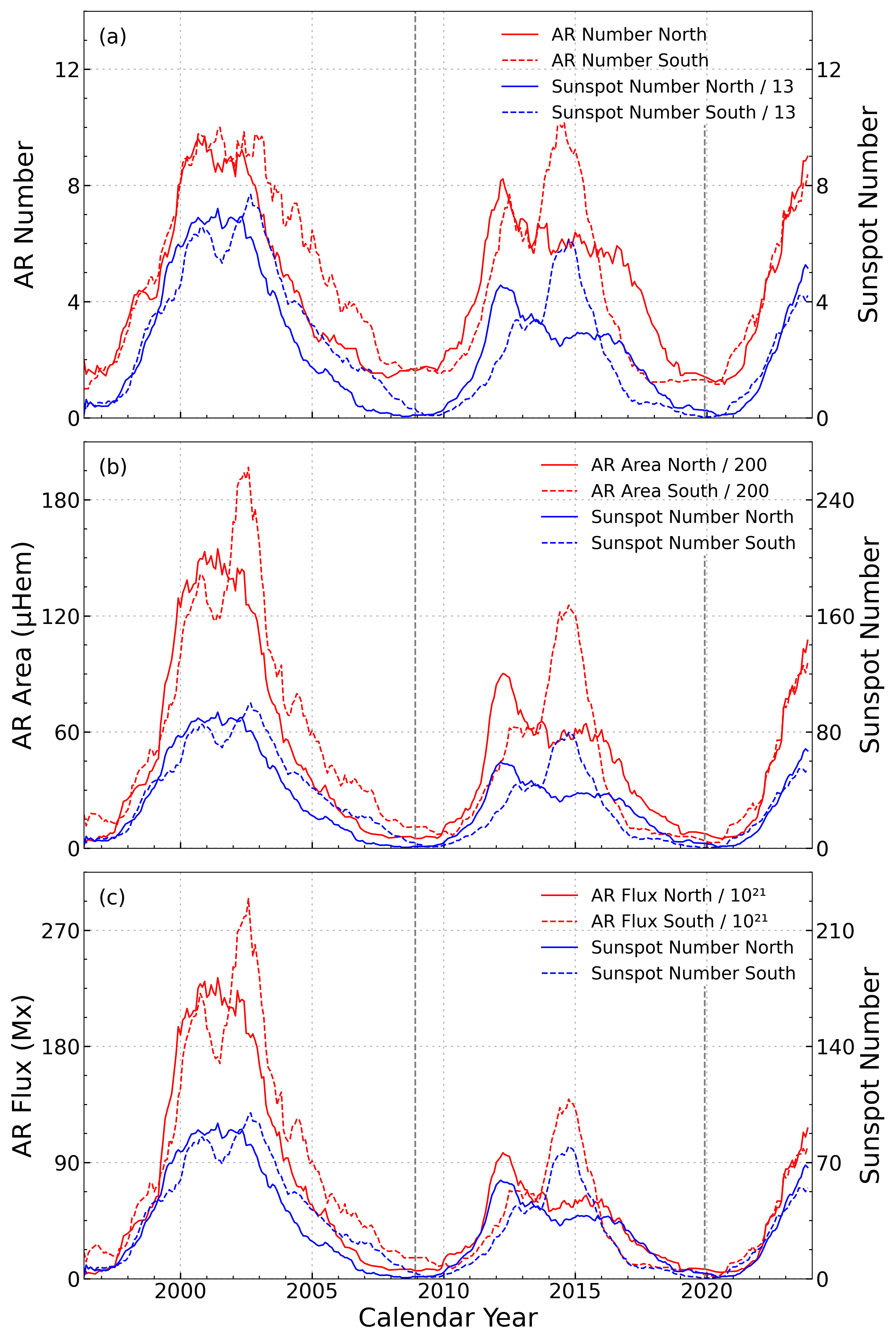}
  \caption{Monthly distribution of AR parameters (red lines) and sunspot numbers (blue lines), where solid lines represent the northern hemisphere and dashed lines represent the southern hemisphere. The gray dashed lines represent the minimum times of the solar cycle. All data are processed with a 13-month smoothing filter to remove high-frequency fluctuations. 
  \label{fig:AR-SSN}}
\end{minipage}
\hfill
\begin{minipage}[t]{0.49\textwidth}
  \centering
  \includegraphics[width=\textwidth, height=12.5cm]{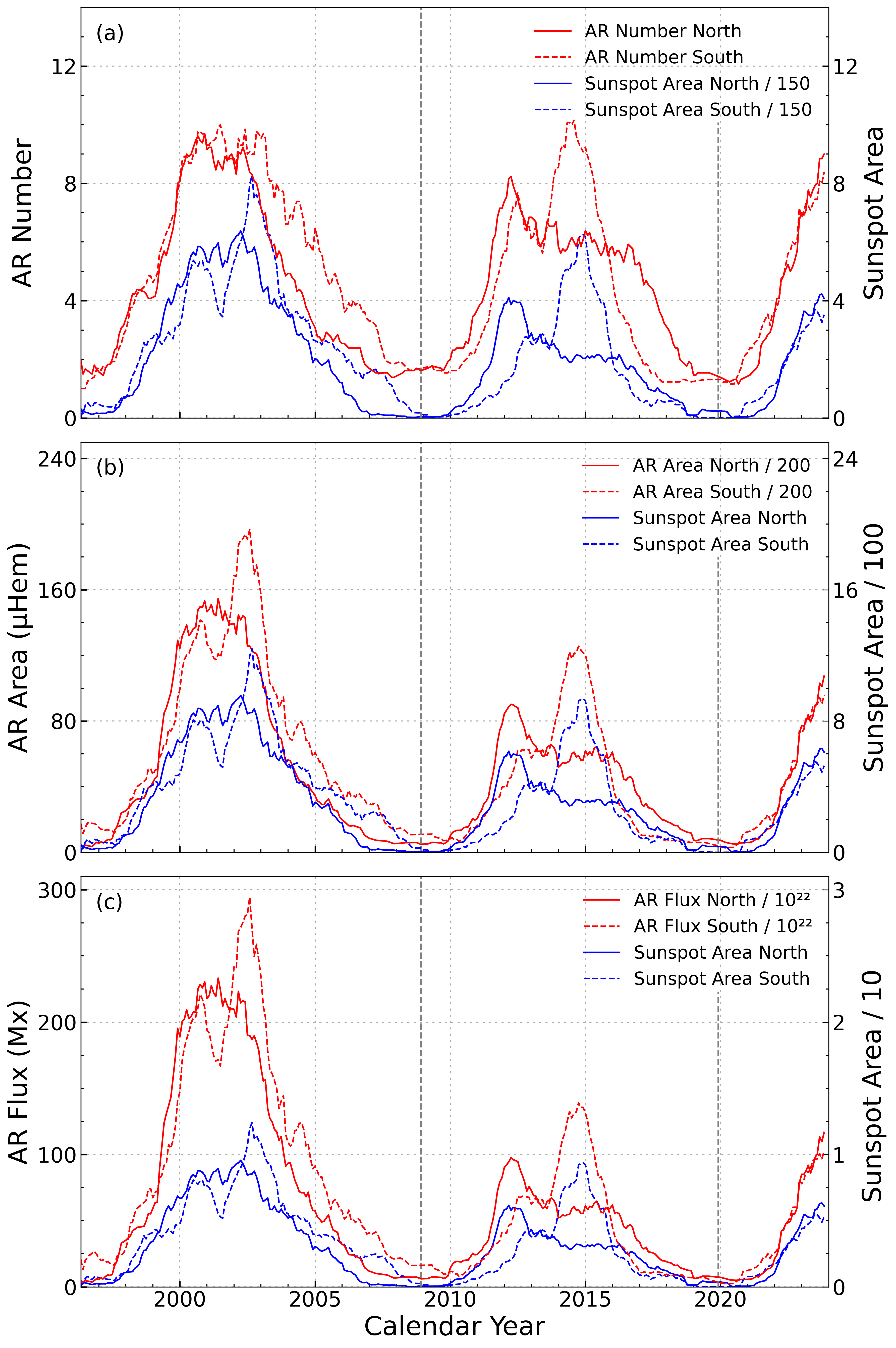}
  \caption{Similar to Figure \ref{fig:AR-SSN}, but showing the comparison between AR parameters and sunspot areas.
  \label{fig:AR-SA}}
\end{minipage}
\end{figure}

 To better investigate the relationship between AR parameters and sunspot activity in different hemispheres, the monthly distributions of AR parameters and sunspot numbers are plotted as shown in Figure \ref{fig:AR-SSN}. Panel (a) shows the monthly distribution of AR number and sunspot numbers in different hemispheres, while panels (b) and (c) correspond to AR area and flux versus sunspot numbers, respectively. Figure \ref{fig:AR-SA} is similar to Figure \ref{fig:AR-SSN}, but shows the monthly distributions of the AR parameters compared to the sunspot areas. Here, some data are scaled to facilitate the visualization of correlation variations (e.g. in panel (a) of Figure \ref{fig:AR-SSN}, the sunspot numbers are divided by 13 in the two hemispheres).

To provide a clearer view of the correlation, Table \ref{tab:sunspot_AR} presents the correlation coefficients between sunspot numbers and areas with various AR parameters in different cycles and hemispheres. Our research shows that there is a strong correlation between the three parameters of AR and sunspot activity across the different solar cycles and hemispheres. As illustrated in panels (b) and (c) of Figure \ref{fig:AR-SA}, the trends of the AR area and flux are similar to those of sunspot areas. This indicates that the formation and evolution of ARs are closely related to the variations in sunspot areas, making the AR area and flux important indicators of sunspot activity intensity. In contrast, the correlation between AR area, flux and sunspot numbers is slightly weaker in the southern hemisphere during SC23 and in the northern hemisphere during SC24.

In addition, there are differences in the correlation between the two hemispheres across the different solar cycles. The correlation in the northern hemisphere is stronger than in the southern one during SC23. In contrast, the southern hemisphere exhibits a higher overall correlation compared to the northern hemisphere in SC24, except for the number of AR and sunspot parameters. This indicates that there are differences in the structure and distribution of ARs, with magnetic reconnection and the evolutionary processes of ARs causing these differences. Furthermore, the solar magnetic field structure becomes more complex and unstable during the cycle minimum, which may lead to discrepancies in the intensity of activity between the two hemispheres. The weak correlation may reflect this complexity.

In conclusion, sunspot areas demonstrate a strong correlation with AR parameters, especially the AR area and flux, which can serve as important indicators of solar magnetic activity and AR evolution. In contrast, while sunspot numbers reflect the frequency of AR formation (see Figure \ref{fig:AR-SSN} (a)), they are less precise in capturing intensity and flux characteristics. The relationship between sunspot activity and AR parameters not only exhibits asymmetry across the different solar cycles, but also shows significant hemispheric differences, highlighting the regionality and complexity of solar magnetic activity \citep{1968SoPh....4..142B,2017ApJ...834...56T}. This provides important reference value for understanding the solar interior magnetic field activity and dynamic processes.

\begin{deluxetable*}{lcccccc}
\tabletypesize{\scriptsize}
\tablewidth{0pt}
\setlength{\tabcolsep}{12pt}
\renewcommand{\arraystretch}{1.6}
\tablecaption{Correlation between AR parameters and sunspot parameters in different cycles and hemispheres. 
\label{tab:sunspot_AR}}
\tablehead{
\colhead{\multirow{1}{*}[-0.4cm]{\centering Sunspot Parameters}} & \colhead{\multirow{1}{*}[-0.4cm]{\centering Solar Cycle}} & \colhead{\multirow{1}{*}[-0.4cm]{\centering Hemisphere}} &
\colhead{AR Number} & \colhead{AR Area \text{($\mu$Hem)}} & \colhead{AR Flux \text{({Mx})}} \\
\cmidrule(r){4-4} 
\cmidrule(r){5-5} 
\cmidrule(r){6-6}
 & & & (r) \hspace{0.8cm} (k) & (r) \hspace{0.8cm} (k) & (r) \hspace{0.8cm} (k) 
} 
\startdata
Sunspot Number & SC23 & N & 0.9888 \hspace{0.5cm} 0.8786 & 0.9099 \hspace{0.5cm} 0.6200 & 0.9100 \hspace{0.5cm} 0.4081 \\
& & S & 0.9738 \hspace{0.5cm} 0.7572 & 0.8888 \hspace{0.5cm} 0.5399 & 0.8965 \hspace{0.5cm} 0.3640 \\
& SC24 & N & 0.9770 \hspace{0.5cm} 0.6213 & 0.8492 \hspace{0.5cm} 0.6825 & 0.8490 \hspace{0.5cm} 0.6617 \\
& & S & 0.9631 \hspace{0.5cm} 0.6089 & 0.9024 \hspace{0.5cm} 0.6637 & 0.9012 \hspace{0.5cm} 0.6095 \\
\midrule
Sunspot Area & SC23 & N & 0.9664 \hspace{0.5cm} 0.7408 & 0.9695 \hspace{0.5cm} 0.0589 & 0.9777 \hspace{0.5cm} 3.9092 \\
& & S & 0.9121 \hspace{0.5cm} 0.6499 & 0.9489 \hspace{0.5cm} 0.0541 & 0.9577 \hspace{0.5cm} 3.6555 \\
& SC24 & N & 0.9446 \hspace{0.5cm} 0.5260 & 0.9744 \hspace{0.5cm} 0.0641 & 0.9798 \hspace{0.5cm} 6.2530 \\
& & S & 0.9144 \hspace{0.5cm} 0.5670 & 0.9803 \hspace{0.5cm} 0.0708 & 0.9837 \hspace{0.5cm} 6.5355 \\
\enddata
\tablecomments{The correlation coefficient (r) represents the Pearson correlation coefficient between sunspot parameters and AR parameters. The values (k) represent the fitting slope obtained from linear regression between sunspot parameters and AR parameters. All the correlation coefficients are above the 99\% confidence levels.}
\end{deluxetable*}

\subsection{Dominant Hemisphere of AR Parameters in Each Year}
\label{Dominant}
Table \ref{tab:numbers} shows the annual total numbers of ARs in each hemisphere. The normalized asymmetry index and the corresponding probabilities for each year are calculated using Equation~\eqref{eq:1} and~\eqref{eq:5}, respectively. A binomial distribution is applied to determine the dominant hemisphere for each year, where the symbol ‘-’ indicates that the levels of ARs in both hemispheres are approximately equal. Tables \ref{tab:areas} and \ref{tab:fluxs} are similar to Table \ref{tab:numbers}, but correspond to AR area and flux, respectively.

\begin{deluxetable*}{lcccccc}
\tabletypesize{\scriptsize}
\tablewidth{0pt} 
\setlength{\tabcolsep}{16pt} 
\renewcommand{\arraystretch}{1.6}
\tablecaption{Asymmetrical distribution of AR number for each year 1996–2023.
\label{tab:numbers}}
\tablehead{
\colhead{Year} & \colhead{$\textit{V}_{N}$} & \colhead{$\textit{V}_{S}$} & \colhead{Asymmetry} & \colhead{Probability} & \colhead{Dominant Hemisphere}
} 
\startdata
1996 & 10 & 10 & 0 & $5.881 \times 10^{-1}$ & - \\
1997 & 40 & 34 & 0.081 & $7.920 \times 10^{-1}$ & - \\
1998 & 37 & 45 & -0.098 & $1.601 \times 10^{-1}$ & - \\
1999 & 98 & 98 & 0 & $5.285 \times 10^{-1}$ & - \\
2000 & 110 & 110 & 0 & $5.269 \times 10^{-1}$ & - \\
2001 & 105 & 109 & -0.019 & $3.663 \times 10^{-1}$ & - \\
2002 & 91 & 118 & -0.129 & $2.625 \times 10^{-2}$ & S \\
2003 & 59 & 79 & -0.145 & $3.673 \times 10^{-2}$ & S \\
2004 & 37 & 72 & -0.321 & $2.525 \times 10^{-4}$ & S \\
2005 & 31 & 53 & -0.262 & $5.817 \times 10^{-3}$ & S \\
2006 & 10 & 41 & -0.608 & $1.694 \times 10^{-6}$ & S \\
2007 & 8 & 17 & -0.360 & $2.164 \times 10^{-2}$ & S \\
2008 & 9 & 6 & 0.200 & $8.491 \times 10^{-1}$ & - \\
2009 & 12 & 7 & 0.263 & $9.165 \times 10^{-1}$ & N \\
2010 & 43 & 26 & 0.246 & $9.853 \times 10^{-1}$ & N \\
2011 & 89 & 69 & 0.127 & $9.528 \times 10^{-1}$ & N \\
2012 & 81 & 78 & 0.019 & $6.244 \times 10^{-1}$ & - \\
2013 & 72 & 102 & -0.172 & $9.255 \times 10^{-3}$ & S \\
2014 & 72 & 109 & -0.204 & $2.300 \times 10^{-3}$ & S \\
2015 & 67 & 59 & 0.064 & $7.886 \times 10^{-1}$ & - \\
2016 & 62 & 27 & 0.393 & $9.999 \times 10^{-1}$ & N \\
2017 & 35 & 12 & 0.489 & $9.998 \times 10^{-1}$ & N \\
2018 & 15 & 9 & 0.250 & $9.242 \times 10^{-1}$ & N \\
2019 & 7 & 5 & 0.167 & $8.062 \times 10^{-1}$ & - \\
2020 & 10 & 22 & -0.375 & $1.003 \times 10^{-2}$ & S \\
2021 & 36 & 43 & -0.089 & $1.841 \times 10^{-1}$ & - \\
2022 & 83 & 82 & 0.006 & $5.618 \times 10^{-1}$ & - \\
2023 & 94 & 87 & 0.0389 & $7.239 \times 10^{-1}$ & - \\
\enddata
\tablecomments{$\textit{V}_{N}$ and $\textit{V}_{S}$ represent the values of AR number in a certain year for the northern and southern hemispheres, respectively. Asymmetry represents \textit{NAI} (see Equation~\eqref{eq:1}).}
\end{deluxetable*}

\begin{deluxetable*}{lcccccc}   
\tabletypesize{\scriptsize}
\tablewidth{0pt}  
\setlength{\tabcolsep}{11pt}  
\renewcommand{\arraystretch}{1.6}  
\tablecaption{Asymmetrical distribution of AR area for each year 1996–2023.
\label{tab:areas}}
\tablehead{
\colhead{Year} & \colhead{$\textit{V}_{N}$ ($\mu$Hem)} & \colhead{$\textit{V}_{S}$ ($\mu$Hem)} & \colhead{Asymmetry} & \colhead{Probability} & \colhead{Dominant Hemisphere}
} 
\startdata
1996 & 7349 & 20135 & -0.465 & $2.962 \times 10^{-3}$ & S \\ 
1997 & 61494 & 45316 & 0.152 & $9.509 \times 10^{-1}$ & N \\ 
1998 & 77117 & 101679 & -0.137 & $2.584 \times 10^{-2}$ & S \\ 
1999 & 304493 & 242349 & 0.114 & $9.965 \times 10^{-1}$ & N \\ 
2000 & 352941 & 306098 & 0.071 & $9.693 \times 10^{-1}$ & N \\ 
2001 & 323792 & 377736 & -0.077 & $1.892 \times 10^{-2}$ & S \\ 
2002 & 277032 & 388023 & -0.167 & $6.714 \times 10^{-6}$ & S \\ 
2003 & 136706 & 173384 & -0.118 & $1.772 \times 10^{-2}$ & S \\ 
2004 & 82764 & 144874 & -0.273 & $1.346 \times 10^{-5}$ & S \\ 
2005 & 48178 & 92102 & -0.313 & $6.281 \times 10^{-5}$ & S \\ 
2006 & 7757 & 74117 & -0.811 & $6.661 \times 10^{-16}$ & S \\ 
2007 & 6078 & 25129 & -0.611 & $9.610 \times 10^{-5}$ & S \\ 
2008 & 4917 & 6273 & -0.121 & $2.744 \times 10^{-1}$ & - \\ 
2009 & 11728 & 6385 & 0.295 & $9.519 \times 10^{-1}$ & N \\ 
2010 & 48001 & 35503 & 0.150 & $9.381 \times 10^{-1}$ & N \\ 
2011 & 201282 & 101226 & 0.331 & $\approx 1.0$ & N \\ 
2012 & 169369 & 147466 & 0.069 & $9.022 \times 10^{-1}$ & N  \\ 
2013 & 125907 & 199266 & -0.226 & $1.880 \times 10^{-5}$ & S \\ 
2014 & 139795 & 291050 & -0.351 & $6.800 \times 10^{-14}$ & S \\ 
2015 & 141957 & 108418 & 0.134 & $9.867 \times 10^{-1}$ & N  \\ 
2016 & 82243 & 28863 & 0.480 & $\approx 1.0$ & N  \\ 
2017 & 45872 & 15897 & 0.485 & $\approx 1.0$ & N  \\
2018 & 14215 & 5434 & 0.447 & $9.904 \times 10^{-1}$ & N  \\ 
2019 & 8168 & 1390 & 0.709 & $9.980 \times 10^{-1}$ & N  \\  
2020 & 6828 & 29887 & -0.628 & $2.063 \times 10^{-5}$ & S \\ 
2021 & 38847 & 56092 & -0.182 & $3.211 \times 10^{-2}$ & S \\ 
2022 & 182123 & 173986 & 0.023 & $6.833 \times 10^{-1}$ & - \\ 
2023 & 218907 & 190829 & 0.069 & $9.240 \times 10^{-1}$ & N \\ 
\enddata
\tablecomments{$\textit{V}_{N}$ and $\textit{V}_{S}$  represent the values of AR area in a certain year for the northern and southern hemispheres, respectively. Asymmetry represents $NAI$ (see Equation~\eqref{eq:1}). For the calculation of the actual distribution probabilities, the AR area are uniformly rounded to an integer and divided by $10^3$ $\mu$Hem.}
\end{deluxetable*}

\begin{deluxetable*}{lcccccc}   
\tabletypesize{\scriptsize}
\tablewidth{0pt}  
\setlength{\tabcolsep}{10.5pt}  
\renewcommand{\arraystretch}{1.6}  
\tablecaption{Asymmetrical distribution of AR flux for each year 1996–2023.
\label{tab:fluxs}}
\tablehead{
\colhead{Year} & \colhead{$\textit{V}_{N}$ (Mx)} & \colhead{$\textit{V}_{S}$ (Mx)} & \colhead{Asymmetry} & \colhead{Probability} & \colhead{Dominant Hemisphere}
} 
\startdata
1996 & $3.707 \times 10^{22}$ & $1.479 \times 10^{23}$ & -0.599 & $2.213 \times 10^{-3}$ & S \\ 
1997 & $4.305 \times 10^{23}$ & $3.193 \times 10^{23}$ & 0.148 & $9.173 \times 10^{-1}$ & N \\ 
1998 & $5.095 \times 10^{23}$ & $7.277 \times 10^{23}$ & -0.176 & $1.922 \times 10^{-2}$ & S \\ 
1999 & $2.328 \times 10^{24}$ & $1.820 \times 10^{24}$ & 0.122 & $9.947 \times 10^{-1}$ & N \\ 
2000 & $2.650 \times 10^{24}$ & $2.362 \times 10^{24}$ & 0.057 & $9.100 \times 10^{-1}$ & N \\ 
2001 & $2.453 \times 10^{24}$ & $2.752 \times 10^{24}$ & -0.057 & $8.697 \times 10^{-2}$ & S \\ 
2002 & $2.109 \times 10^{24}$ & $2.920 \times 10^{24}$ & -0.161 & $1.243 \times 10^{-4}$ & S \\ 
2003 & $1.146 \times 10^{24}$ & $1.322 \times 10^{24}$ & -0.071 & $1.260 \times 10^{-1}$ & - \\ 
2004 & $6.543 \times 10^{23}$ & $1.092 \times 10^{24}$ & -0.251 & $3.026 \times 10^{-4}$ & S \\ 
2005 & $4.012 \times 10^{23}$ & $7.247 \times 10^{23}$ & -0.287 & $8.473 \times 10^{-4}$ & S \\ 
2006 & $4.800 \times 10^{22}$ & $5.051 \times 10^{23}$ & -0.826 & $5.504 \times 10^{-12}$ & S \\ 
2007 & $3.984 \times 10^{22}$ & $1.926 \times 10^{23}$ & -0.657 & $2.441 \times 10^{-4}$ & S \\ 
2008 & $2.790 \times 10^{22}$ & $4.319 \times 10^{22}$ & -0.215 & $2.266 \times 10^{-1}$ & - \\ 
2009 & $7.256 \times 10^{22}$ & $3.929 \times 10^{22}$ & 0.297 & $8.867 \times 10^{-1}$ & - \\ 
2010 & $2.941 \times 10^{23}$ & $1.814 \times 10^{23}$ & 0.237 & $9.605 \times 10^{-1}$ & N \\ 
2011 & $1.066 \times 10^{24}$ & $5.233 \times 10^{23}$ & 0.342 & $\approx 1.0$ & N \\ 
2012 & $9.039 \times 10^{23}$ & $8.004 \times 10^{23}$ & 0.061 & $8.005 \times 10^{-1}$ & - \\ 
2013 & $6.319 \times 10^{23}$ & $1.064 \times 10^{24}$ & -0.255 & $3.339 \times 10^{-4}$ & S \\ 
2014 & $6.876 \times 10^{23}$ & $1.605 \times 10^{24}$ & -0.400 & $2.505 \times 10^{-10}$ & S \\ 
2015 & $7.248 \times 10^{23}$ & $5.708 \times 10^{23}$ & 0.119 & $9.207 \times 10^{-1}$ & N \\ 
2016 & $4.130 \times 10^{23}$ & $1.522 \times 10^{23}$ & 0.461 & $9.999 \times 10^{-1}$ & N \\ 
2017 & $2.339 \times 10^{23}$ & $9.565 \times 10^{22}$ & 0.420 & $9.932 \times 10^{-1}$ & N \\
2018 & $6.157 \times 10^{22}$ & $2.663 \times 10^{22}$ & 0.396 & $9.102 \times 10^{-1}$ & N \\ 
2019 & $4.495 \times 10^{22}$ & $5.265 \times 10^{21}$ & 0.790 & $3.125 \times 10^{-2}$ & S \\  
2020 & $2.722 \times 10^{22}$ & $1.514 \times 10^{23}$ & -0.695 & $6.561 \times 10^{-4}$ & S \\ 
2021 & $1.894 \times 10^{23}$ & $3.074 \times 10^{23}$ & -0.238 & $3.245 \times 10^{-2}$ & S \\ 
2022 & $9.944 \times 10^{23}$ & $9.504 \times 10^{23}$ & 0.023 & $6.401 \times 10^{-1}$ & - \\ 
2023 & $1.187 \times 10^{24}$ & $1.011 \times 10^{24}$ & 0.080 & $\approx 8.999 \times 10^{-1}$ & - \\ 
\enddata
\tablecomments{$\textit{V}_{N}$ and $\textit{V}_{S}$ represent the values of AR flux in a certain year for the northern and southern hemispheres, respectively. Asymmetry represents $NAI$ (see Equation~\eqref{eq:1}). For the calculation of the actual distribution probabilities, the AR flux are uniformly divided by $10^{22}$ Mx.}
\end{deluxetable*}

As shown in Table \ref{tab:numbers}, one can see that, among the hemispheric asymmetry of the AR number within 28 years (from 1996 to 2023), the hemispheric asymmetry is highly significant in 15 cases and insignificant in 13 cases. This indicates that more than 53\% of the observations exhibit a pronounced N-S asymmetry. This finding suggests that the hemispheric asymmetry in AR number is a genuine physical phenomenon during this period. We observe that there is no dominant hemisphere in terms of AR number during the ascending phase of SC23, whereas the southern hemisphere shows dominance after 2002. The northern hemisphere dominates before and after the maximum time of SC24, while the southern hemisphere dominates during the cycle maximum. For the ongoing SC25, as the data only cover the ascending phase, the trend is likely similar to the SC23 with no clear dominant hemisphere, except for the year 2020.

From Table \ref{tab:areas}, it is evident that the hemispheric distribution of the AR area shows a significant asymmetry in 26 cases (from 1996 to 2023), with only 2 cases being insignificant. This means that more than 90\% of the observations show a pronounced N-S asymmetry, indicating that the hemispheric asymmetry of the AR area during this period is a genuine physical phenomenon. Observations reveal that the dominant hemisphere for the AR area was inconsistent before 2001; however, the southern hemisphere consistently dominates during SC23 (after 2001), except for 2008. In addition, the southern hemisphere dominates during the maximum time of SC24, while the northern hemisphere is dominant before and after this period. 

From Table \ref{tab:fluxs}, the hemispheric distribution of the AR flux reveals a significant asymmetry in 22 cases (from 1996 to 2023), with only 6 cases being insignificant. This indicates that more than 78\% of the observations exhibit a pronounced N-S asymmetry, confirming that the hemispheric asymmetry for AR flux during this period is a genuine physical phenomenon. The overall trend in AR flux closely resembles that AR area throughout the entire period, with the exception of certain years.

The magnetic field in the solar polar regions plays a crucial role throughout the solar cycle, gradually accumulating and reconfiguring during the solar polarity reversal. The polarity reversal period acts as a transition phase, leading to differences in the dominant hemisphere before and after the reversal time. Significant differences are observed across the different solar cycles: the southern hemisphere's magnetic field appears to be stronger during SC23. In contrast, AR parameters are dominant in the northern hemisphere before and after the maximum time of SC24, while the southern hemisphere leads during the cycle maximum (2013–2014). During SC25, the trend of the AR number likely resembles that of SC23, while the AR area and flux are likely to be the opposite of SC24.

\subsection{Dominant Hemisphere for Each SC23–SC25}
\label{subsec:Domin}
To determine the dominant hemisphere for the AR parameters in each solar cycle, we calculated the counts of AR number, area, and flux in each hemisphere in different solar cycles. Table \ref{tab:all} provides the specific numerical values. Here, the time interval of SC23 is from 1996 May to 2008 November, SC24 is from 2008 December to 2019 November, and SC25 is from 2019 December to 2023 November.

\begin{deluxetable*}{lcccccc}   
\tabletypesize{\scriptsize}
\tablewidth{0pt}  
\setlength{\tabcolsep}{8pt}  
\renewcommand{\arraystretch}{1.6}  
\tablecaption{Asymmetrical distribution of AR parameters during SC23 to SC25.}
\label{tab:all}
\tablehead{
\colhead{AR Parameters} & \colhead{Solar Cycle} & \colhead{$\textit{V}_{N}$} & \colhead{$\textit{V}_{S}$} & \colhead{Asymmetry} & \colhead{Probability} & \colhead{Dominant Hemisphere}
} 
\startdata
{\centering AR Number} 
    & SC23 & 644 & 792 & -0.103 & $4.155 \times 10^{-5}$ & S \\ 
    & SC24 & 556 & 502 & 0.051 & $9.546 \times 10^{-1}$ & N \\ 
    & SC25 & 223 & 235 & -0.026 & $2.718 \times 10^{-1}$ & - \\ \midrule
{\centering AR Area ($\mu$Hem)}
    & SC23 & 1689849 & 1997215 & -0.083 & $1.935 \times 10^{-7}$ & S \\ 
    & SC24 & 989305 & 940528 & 0.025 & $8.677 \times 10^{-1}$ & - \\ 
    & SC25 & 446704 & 451164 & -0.005 & $4.337 \times 10^{-1}$ & - \\ \midrule
{\centering AR Flux (Mx)}
    & SC23 & $1.283 \times 10^{25}$ & $1.493 \times 10^{25}$ & -0.076 & $3.081 \times 10^{-5}$ & S \\ 
    & SC24 & $5.139 \times 10^{24}$ & $5.062 \times 10^{24}$ & 0.008 & $6.109 \times 10^{-1}$ & - \\ 
    & SC25 & $2.398 \times 10^{24}$ & $2.422 \times 10^{24}$ & -0.005 & $4.457 \times 10^{-1}$ & - \\ 
\enddata
\tablecomments{$\textit{V}_{N}$ and $\textit{V}_{S}$ represent the values of three AR parameters in a certain year for the northern and southern hemispheres, respectively. Asymmetry represents $NAI$ (see Equation~\eqref{eq:1}). For the calculation of the actual distribution probabilities, the AR area are uniformly rounded to an integer and divided by $10^3$ $\mu$Hem, and the AR flux are uniformly divided by $10^{22}$ Mx.}
\end{deluxetable*}

There is a significant difference in the dominant hemisphere between different solar cycles. The three AR parameters dominate in the southern hemisphere during SC23, while the activity levels of the three AR parameters are nearly equal in the two hemispheres during SC25. In SC24, the AR number highlights the dominance of the northern hemisphere, while the AR area and flux exhibit comparable intensities in both hemispheres.

As shown in Table \ref{tab:all}, we can easily see that the dominant hemisphere varies among the three AR parameters. The dominant hemispheres for AR area and flux are consistent across the different solar cycles, whereas they differ from the AR number. Specifically, the northern hemisphere dominates for AR number during SC24, while AR area and flux show nearly equal intensities in both hemispheres.

From the above analyses, we can conclude that the hemispheric ARs vary in different ways at different solar cycles and parameters.

\subsection{Slope of Asymmetry in AR Parameters}
\label{Asymmetry}
\begin{figure*}[ht!]
\plotone{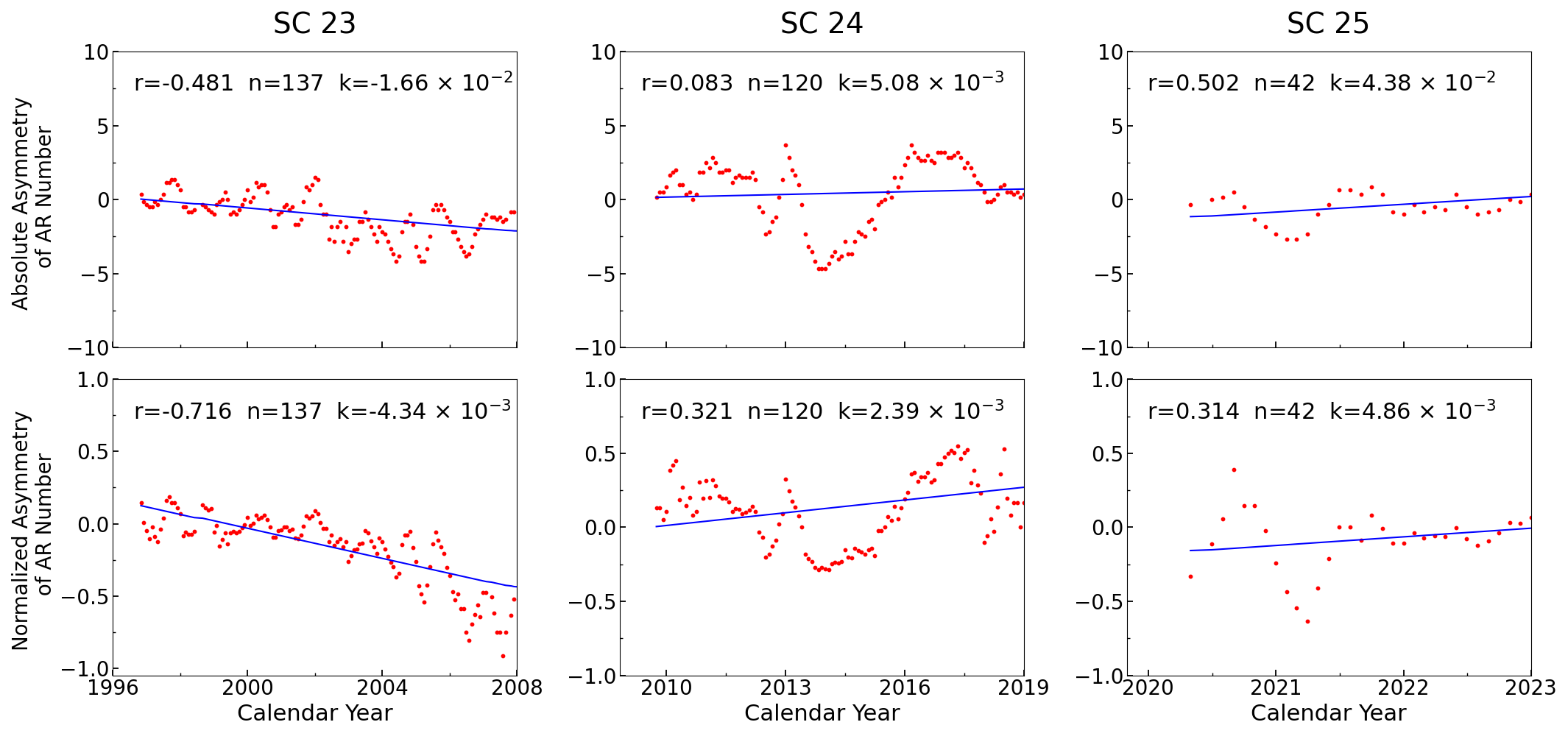}
\caption{Regression fit lines for the montly values (six-month average) of N-S asymmetry in AR number across the different cycles. Upper panel: scatter plot of absolute asymmetry values over time; Lower panel: similar to the upper panel but corresponding to normalized asymmetry. Red circles represent actual data points, and the blue line indicates the fitted regression line. The number of data points (\textit{n}), correlation coefficient (\textit{r}) and fitted slope (\textit{k}) are given on the plots.
\label{fig:number}}
\end{figure*}

\begin{figure*}[ht!]
\plotone{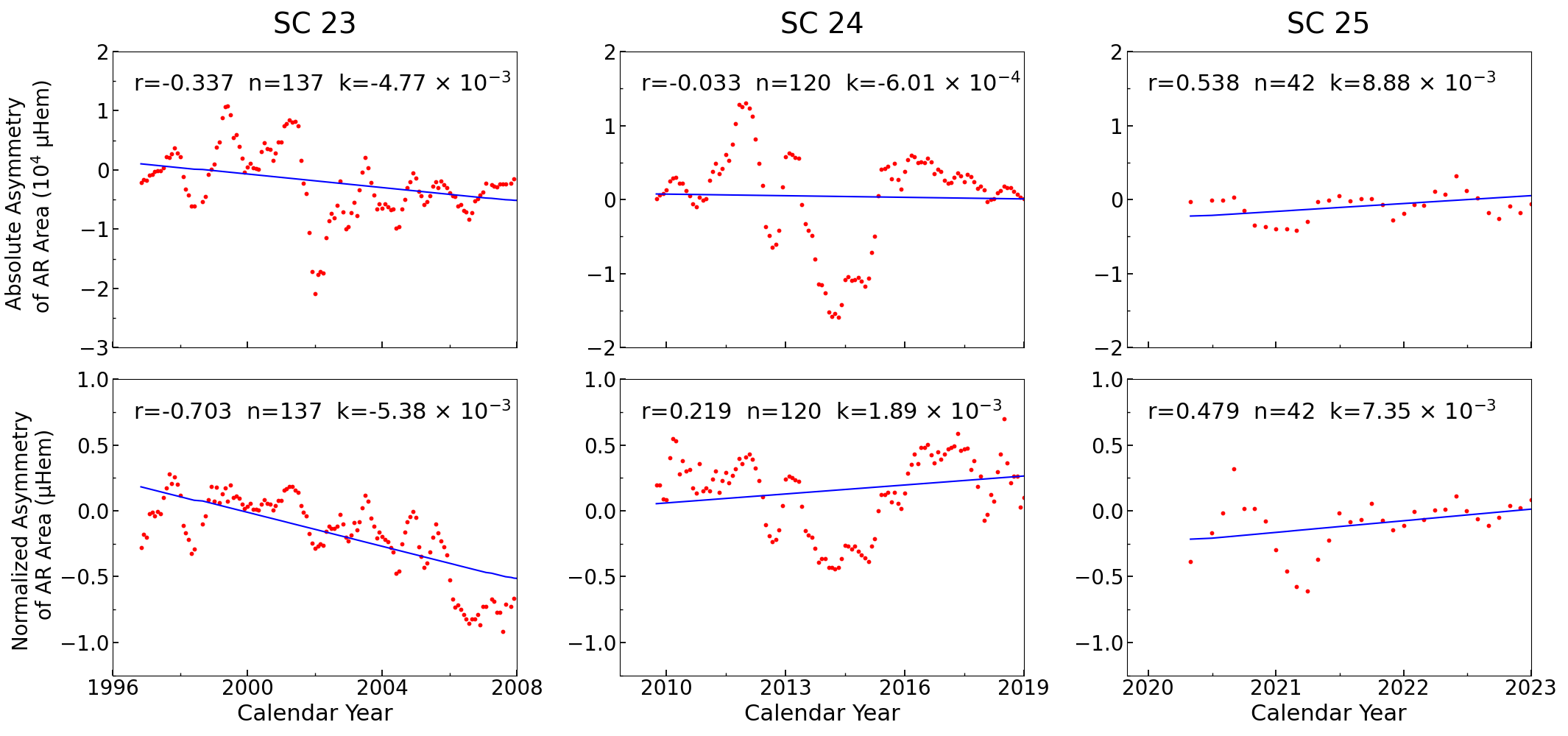}
\caption{Same as Figure \ref{fig:number}, but showing the regression lines of the N-S asymmetry in AR area across the different cycles.
\label{fig:area}}
\end{figure*}

\begin{figure*}[ht!]
\plotone{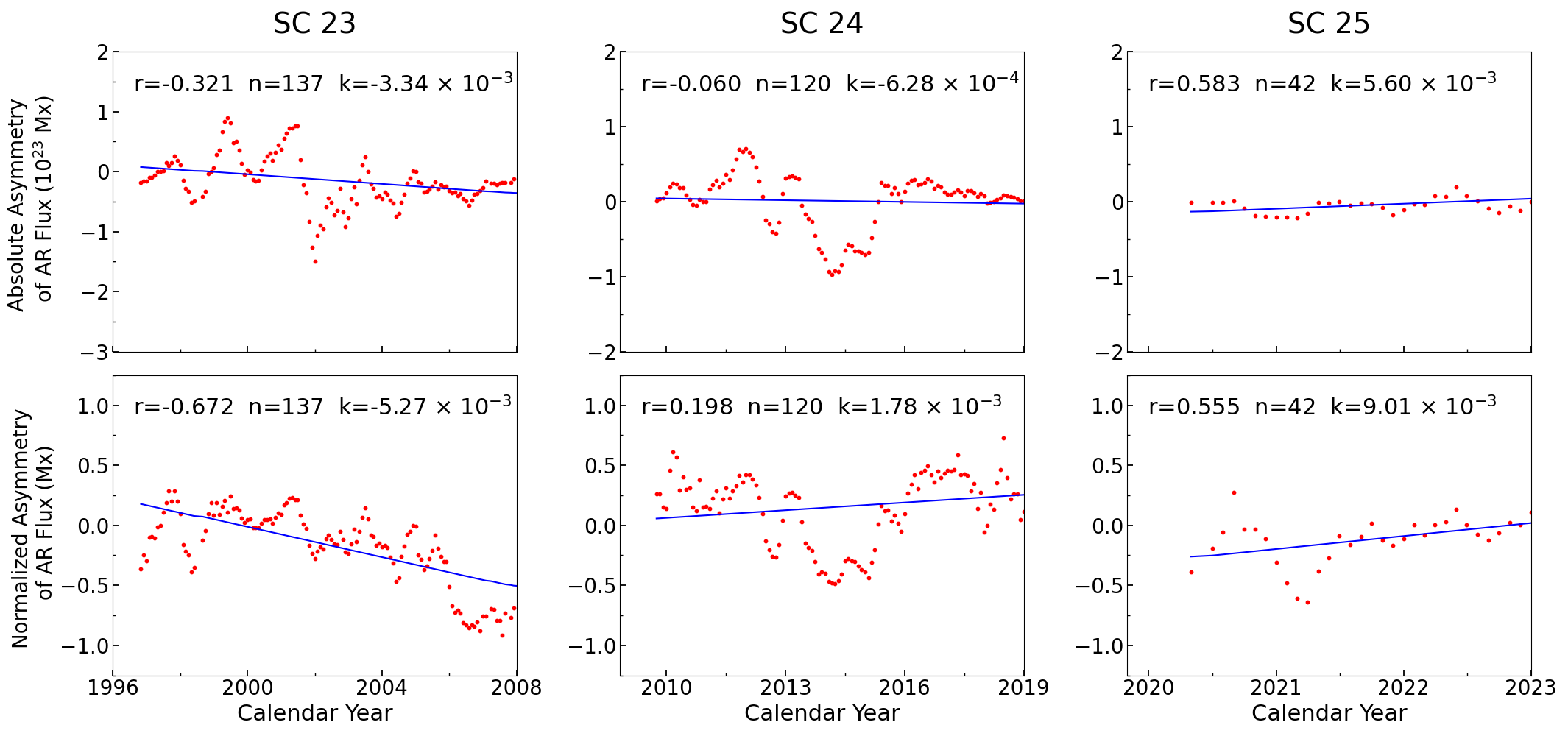}
\caption{Same as Figures \ref{fig:number} and \ref{fig:area}, but showing the regression lines of the N-S asymmetry in AR flux across the different cycles.
\label{fig:flux}}
\end{figure*}

Here, we first use the normalized asymmetry index (see Equation \eqref{eq:1}) and the absolute asymmetry index (see Equation\eqref{eq:2}) to quantify the asymmetry of the three AR parameters in the two hemispheres. Subsequently, we plot the monthly time series of both indices for AR parameters spanning from 1996 May to 2023 November (see Figure \ref{fig:number} to \ref{fig:flux}). Figure \ref{fig:number} shows the time series of the two asymmetry indices of the AR number. We observe that the fitted slopes of both asymmetry indices are negative in SC23. However, the slopes are positive in SC24 and SC25. Figures \ref{fig:area} and \ref{fig:flux} display the time series of the two asymmetry indices for the AR area and the flux, respectively. We observe that the slopes of the two asymmetry indices for the AR area and flux in SC24 are exactly opposite: the fitted slope of the absolute asymmetry index is negative, while the fitted slope of the normalized asymmetry index is positive. This could be due to the low values of the AR parameters during the minimum time of SC24. Then we select data where the monthly total AR counts in both hemispheres are greater than or equal to 5 to analyze the normalized asymmetry index of the three AR parameters. We find that for both the AR area and flux, the fitted slope in SC24 changes from positive to negative. This result further verifies and strengthens the research of \citet{2022MNRAS.514.1140Z}, who found that the fitted slopes of hemispheric sunspot activity during SC20 to SC24 are negative and argued that the correct asymmetry index for regression analysis should be based on the absolute difference in solar activity between the two hemispheres. If one wants to use the normalized asymmetry index, only solar activity above a certain threshold should be considered.

During the solar cycle minimum, the AR number is very low, and a small denominator can result in large values. Therefore, we primarily use the absolute asymmetry index to analyze the fitting slope of the AR parameters. Based on Figure \ref{fig:number}, there is a correlation between asymmetry and time at the confidence level 99\% for the AR number during SC23 (the correlation coefficient is -0.481 and the number of data points is 137). The equation of the regression line is \textit{Asymmetry} = $\left( -1.661 \times 10^{-2} \pm 2.604 \times 10^{-3} \right) \times \textit{Time} + \left( 1.141 \times 10^{-1} \pm 2.604 \times 10^{-3} \right)$, where the unit of time is the year. However, there may be no correlation between asymmetry and time during SC24 (the correlation coefficient is 0.083 and the number of data points is 120) and the equation of the regression line is \textit{Asymmetry} = $\left( 5.076 \times 10^{-3} \pm 5.644 \times 10^{-3} \right) \times \textit{Time} + \left( 1.240 \times 10^{-1} \pm 5.644\times 10^{-3} \right)$. During SC25, there is a correlation between asymmetry and time at 99\% confidence level (the correlation coefficient is 0.502 and the number of data points is 42) and the equation of the regression line is \textit{Asymmetry} = $\left( 4.378 \times 10^{-2} \pm 1.191 \times 10^{-2} \right) \times \textit{Time} + \left(-1.370 \pm 1.191 \times 10^{-2} \right)$.

Based on Figure \ref{fig:area}, there is a correlation between asymmetry and time at 99\% confidence level for the AR area during SC23 (the correlation coefficient is -0.337 and the number of data points is 137). The equation of the regression line is \textit{Asymmetry} = $\left( -4.771 \times 10^{-3} \pm 1.147 \times 10^{-3} \right) \times \textit{Time} + \left( 1.272 \times 10^{-1} \pm 1.147 \times 10^{-3} \right)$, where the unit of time is the year. However, there may be no correlation between asymmetry and time during SC24 (the correlation coefficient is -0.033 and the number of data points is 120) and the equation of the regression line is \textit{Asymmetry} = $\left( -6.014 \times 10^{-4} \pm 1.666 \times 10^{-3} \right) \times \textit{Time} + \left( 7.853 \times 10^{-2} \pm 1.666\times 10^{-3} \right)$. During SC25, there is a correlation between asymmetry and time at 99\% confidence level (the correlation coefficient is 0.538 and the number of data points is 42) and the equation of the regression line is \textit{Asymmetry} = $\left( 8.877 \times 10^{-3} \pm 2.199\times 10^{-3} \right) \times \textit{Time} + \left( -2.669 \times 10^{-1}\pm 2.199 \times 10^{-3} \right)$.

Based on Figure \ref{fig:flux}, there is a correlation between asymmetry and time at 99\% confidence level for the AR flux during SC23 (the correlation coefficient is -0.321 and the number of data points is 137). The equation of the regression line is \textit{Asymmetry} = $\left( -3.344 \times 10^{-3} \pm 8.487 \times 10^{-4} \right) \times \textit{Time} + \left( 9.278 \times 10^{-2} \pm 8.425 \times 10^{-4} \right)$, where the unit of time is the year. However, there may be no correlation between asymmetry and time during SC24 (the correlation coefficient is -0.060 and the number of data points is 120) and the equation of the regression line is \textit{Asymmetry} = $\left( -6.277  \times 10^{-4}\pm 9.595 \times 10^{-4} \right) \times \textit{Time} + \left( 4.623 \times 10^{-2} \pm 9.515\times 10^{-4} \right)$. During SC25, there is a correlation between asymmetry and time at 99\% confidence level (the correlation coefficient is 0.583 and the number of data points is 42) and the equation of the regression line is \textit{Asymmetry} = $\left(5.601 \times 10^{-3} \pm 1.235\times 10^{-3} \right) \times \textit{Time} + \left( -1.624 \times 10^{-1}\pm 1.206 \times 10^{-3} \right)$. 

\begin{figure*}[ht!]
\centering
\includegraphics[width=0.72\textwidth,height=0.42\textheight]{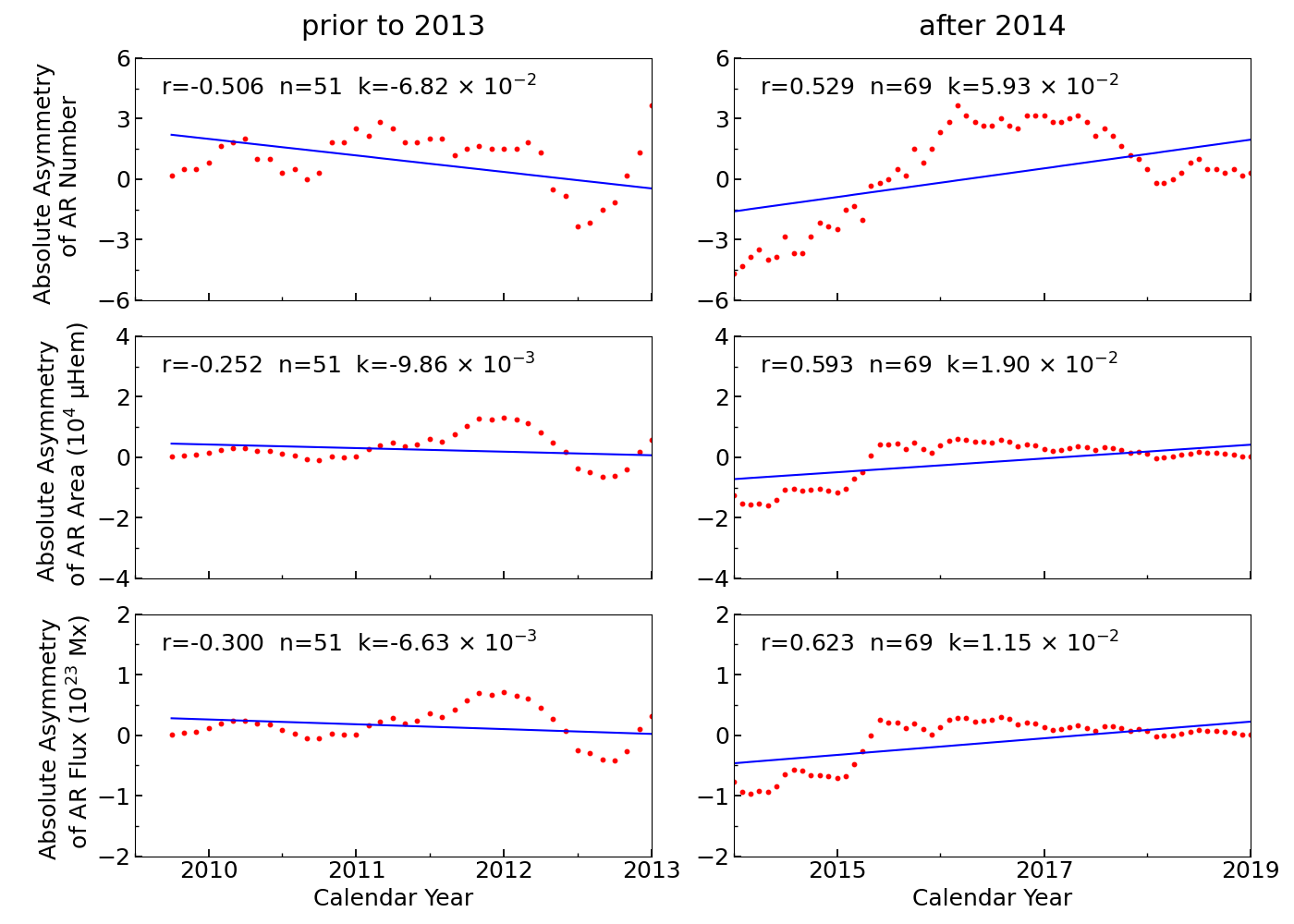}
\caption{Regression fit lines for the montly values (six-month average) of N-S asymmetry in three AR parameters during SC24. Left panel: The relationship between the three AR parameter asymmetry and time prior to 2013. Right panel: similar to the left panel, but corresponding to the period after 2014. Red circles represent actual data points, and the blue line indicates the fitted regression line. The number of data points (\textit{n}), correlation coefficient (\textit{r}) and fitted slope (\textit{k}) are given on the plots.
\label{fig:24}}
\end{figure*}

For SC24, we plot the time variation of the absolute asymmetry index of the three AR parameters before 2013 and after 2014 (see Figure \ref{fig:24}), respectively. Before 2013, the asymmetry of three AR parameters shows a correlation with time at 90\% confidence levels, and their asymmetry fitting slopes are all negative. After 2014, the asymmetry of three AR parameters shows a correlation with time at 99\% confidence levels, and their asymmetry fitting slopes are all positive.

For the three parameters of solar ARs, their fitting trends are consistent with sunspot activity \citep{2022MNRAS.514.1140Z}. In SC23, the absolute asymmetry fitting slopes for the AR parameters are negative, while the fitting slopes are positive in SC25. For SC24, we observe that the fitting slopes for the three AR parameters are negative before 2013, and the fitting slopes become positive after 2014. \citet{2019ApJ...873..121L} mentioned that the sign of the slope changes near the maximum of SC24. This behavior is not seen in other cycles, and they consider it a unique property of SC24.

\subsection{Cumulative Trend of the Hemispheric AR parameters}
\label{Cumulative trend}

\begin{figure*}[ht!]
\centering
\includegraphics[width=0.9\textwidth]{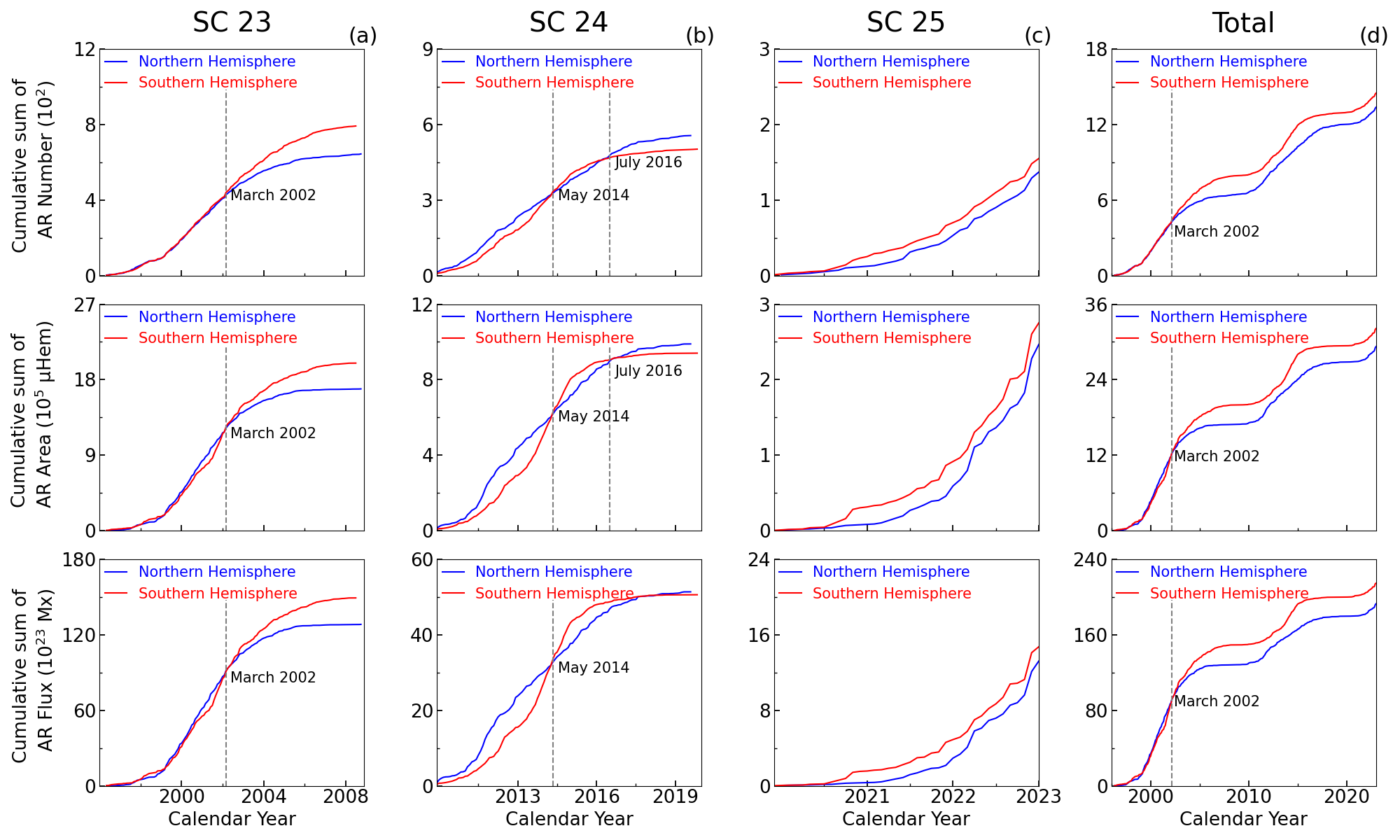}
\caption{Panel (a): cumulative counts of AR parameters in the southern (red) and northern (blue) hemispheres during SC23. Panels (b), (c) and (d) are similar to panel (a), but correspond to the cumulative counts for SC24, SC25, and the entire period, respectively.
\label{fig:cul}}
\end{figure*}

\begin{figure*}[ht!]
\centering
\includegraphics[width=0.9\textwidth]{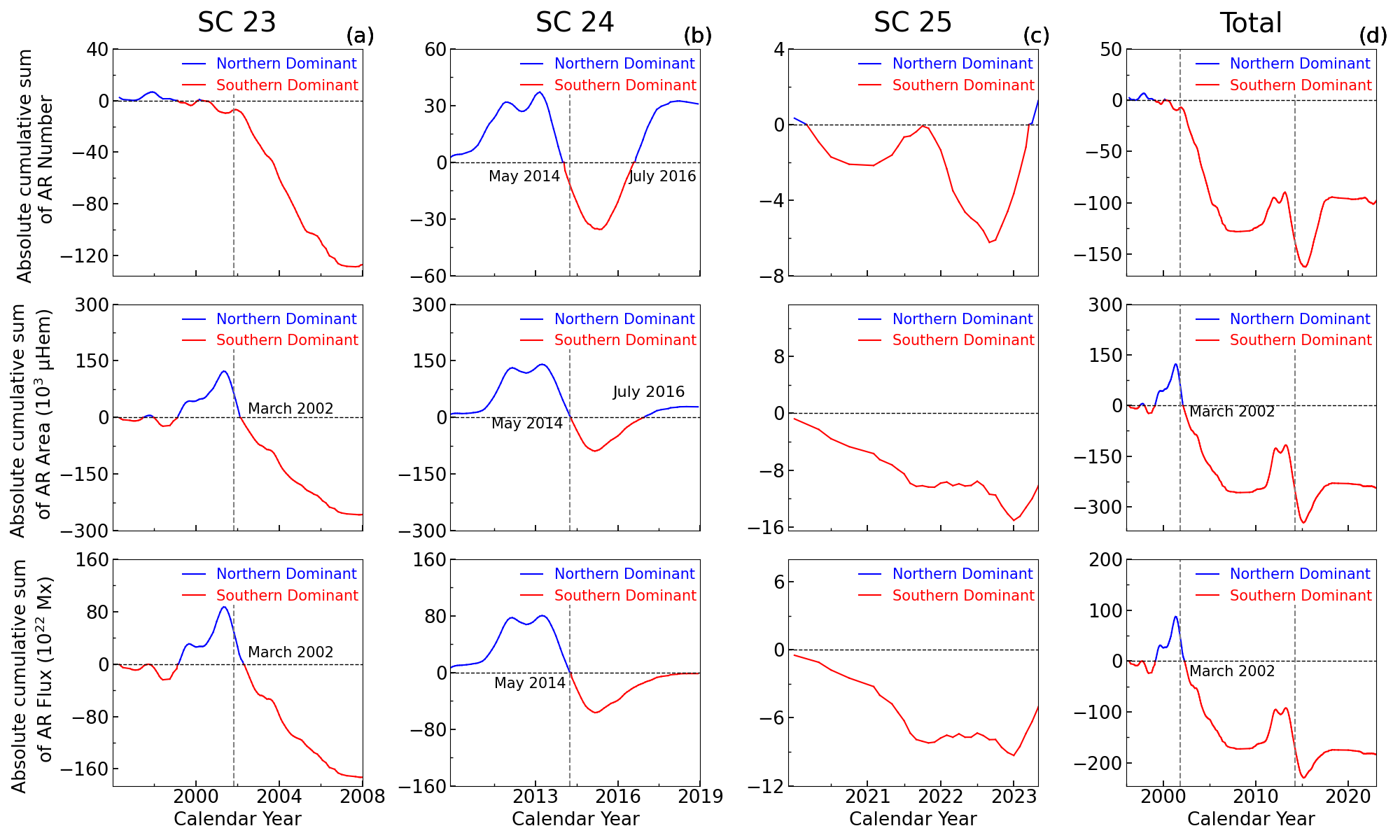} 
\caption{Panel (a): cumulative asymmetry in AR parameters across different hemispheres during SC23. The red and blue lines indicate an excess in the cumulative sum between the southern and northern hemispheres. Panels (b), (c) and (d) are similar to panel (a), corresponding to SC24, SC25, and the entire time period, respectively.
\label{fig:abs}}
\end{figure*}

To better understand the systematic change in hemispheric asymmetry during different phases of a solar cycle, we applied the hemispheric cumulative sum to study the intercycle trend of the dominance of AR parameters between the two hemispheres. Figure \ref{fig:cul} presents the cumulative counts of AR number, area, and flux over time from 1996 May to 2023 November.

Figure \ref{fig:cul}(a) illustrates that the cumulative counts of AR number exhibit distinct temporal behaviors compared to that of AR area and flux during SC23. The AR number shows roughly equivalent activity levels between the two hemispheres prior to 2002 March. For AR area and flux, we can observe that the two lines are very close in the two hemispheres, but overall, the cumulative counts in the northern hemisphere is slightly higher than in the southern one. After 2002 March, the cumulative counts of the three AR parameters have become significantly higher in the southern hemisphere, with the divergence becoming larger and larger between the two hemispheres at the end of the cycle. That is, from a global perspective, the southern hemisphere clearly dominates in all AR parameters during this period.

Figure \ref{fig:cul}(b) shows the hemispheric variations in the cumulative counts of AR parameters during SC24. The cumulative counts for all AR parameters in the northern hemisphere exceed those in the southern one prior to 2014 May; however, the cumulative counts of AR number and area in the southern hemisphere are larger than those in the northern one from 2014 May to 2016 July. Following this period, the northern hemisphere once again leads in AR number and area, with the hemispheric difference continuing to grow at the end of the cycle. In terms of AR flux, the activity in the southern hemisphere is significantly stronger than that in the northern one from 2014 May to 2017 July; the two lines are very close at the end of SC24. Thus, from a global perspective, the cumulative counts of AR number and area are dominated by the northern hemisphere before and after the maximum time of SC24, while three parameters are dominated by the southern hemisphere during the cycle maximum.

Figure \ref{fig:cul}(c) shows the cumulative evolution of the AR parameters across the two hemispheres during SC25. The activity in the southern hemisphere consistently exceeds that in the northern hemisphere at the beginning of the cycle. As the cycle progresses, the divergence between the two hemispheres becomes increasingly pronounced.

Figure \ref{fig:cul}(d) presents the hemispheric evolution of AR parameters from 1996 May to 2023 November. It can be seen that the trends before 2002 March resemble those observed in the early phase of SC23 (before 2002 March). However, the cumulative counts of AR parameters in the southern hemisphere consistently is larger than that in the northern one after 2002 March, and the divergence between the two hemispheres becomes increasingly pronounced over time. These findings suggest that the southern hemisphere exhibits a clear dominance during this period.

To better show the trend difference of the hemispheric cumulative counts, Figure \ref{fig:abs} shows the variation of the cumulative absolute asymmetry (see Equation \eqref{eq:3}) of the AR parameters in different solar cycles. The red and blue lines indicate an excess in the cumulative sum between the southern and northern hemispheres.

Our findings show that the patterns for each solar cycle remain consistent with those presented in Figure \ref{fig:cul}, indicating that the trends of AR area and flux are similar. The temporal behavior of AR area and flux differs from that of AR number during SC23; however, from a global perspective, the southern hemisphere demonstrates clear dominance across all parameters, with hemispheric asymmetry steadily increasing over time. The northern hemisphere dominates in terms of AR number during SC24. In addition, the three AR parameters are dominated by the southern hemisphere during the ascending phase of SC25 and throughout the period.

\section{Conclusions and Discussions}
\label{sec:CD}
For a single synoptic magnetogram, the magnetic field distribution of ARs is primarily described by three parameters: number, area, and flux, which are not considered to be symmetrically distributed in the two hemispheres. This study systematically analyzes the hemispheric distribution of AR parameters based on a homogeneous AR database, covering SC23 to part of SC25.

The distribution of the AR parameters shows significant asymmetry between the two hemispheres, showing the cycle-dependent properties (see Figure \ref{fig:ns}). Our results align with the sunspot velocity trends derived by \citet{2010MNRAS.405.1040L} in SC24. They observed that the drift velocity in the northern hemisphere was higher than in the southern one during the rising phase (2010–2014). After 2015, the velocity increased in both hemispheres, but migration in the southern hemisphere was faster than that in the northern one. \citet{2024ApJS..272....5Z} suggested that the asymmetry of the drift velocity in the two hemispheres for ARs may be related to the unusual migration of the magnetic field in SC24. The latitude distribution of the AR parameters follows the ``Sp\"orer's Law of Zones''. Additionally, compared to SC23, both the AR area and flux in SC24 have significantly decreased, which is consistent with sunspot activity. \citet{1980SoPh...68..141B} found that the activity regions of solar faculae exhibit distinct latitude distribution differences at different stages of the cycle, a phenomenon that is also similar to the latitude distribution of sunspots. We also observe differences in activity intensity and peak timing in both hemispheres, with the northern hemisphere reaching its peak earlier than the southern one. This may originate from the uneven distribution and evolution of the magnetic field \citep{1986SoPh..106...35S,2022SoPh..297...33J}. The observed hemispheric asymmetries may result from complex interactions between dipolar and quadrupolar magnetic field modes during their evolution in different hemispheres \citep{2021ApJ...919...36K,2023AdSpR..71.1984Z}. In addition, \citet{2019MNRAS.488..111D,2020MNRAS.494.4930D} found that the solar quasi-biennial oscillations (QBOs) exhibit hemispheric asymmetry. On the one hand, the QBOs of solar H$ \alpha $ flare activity show phase and amplitude asymmetry in the two hemispheres, reflecting the complex dynamics of solar activity in each hemisphere. Furthermore, although the QBOs of polar faculae exhibits global coherence, its manifestation differs between the hemispheres, revealing the intrinsic complexity of the N-S asymmetry in solar activity. 

The area and flux of AR exhibit a strong positive correlation between different solar cycles (see Figure \ref{fig:cor}), confirming and reinforcing the linear relationship between the AR area and flux \citep{1966ApJ...144..723S}. The correlation between AR area and flux varies across the different cycles (with SC24 being smaller than SC23), and their trends differ from the AR number. The complex magnetic fields of ARs serve as an energy source for intense solar activities such as flares, jets, and CMEs \citep{2006ApJ...644.1273J,2017ApJ...834...56T,2022AdSpR..70.1549T}. Although the AR number is a representative indicator of solar activity, it is insufficient to fully reflect the actual intensity of the ARs. \citet{2020ApJ...900..128L} found a strong negative correlation between the proportion of eruptive to the total number of flares and the flux of AR. This implies that even if the number of AR is high, the total magnetic flux may not increase significantly if most of the regions are small, thereby influencing the eruptive characteristics of flares. The results in Section \ref{subsec:b} align with the previous findings by \citet{2020ApJ...900..128L}, where the variation of the AR number differs from that of the AR area and flux during SC23. Therefore, in future studies of solar activity monitoring and periodic variations, parameters such as area and flux may become more critical and better reflect the essence of ARs.

Compared to sunspot numbers or sunspot group numbers, sunspot areas are considered to have greater physical significance \citep{2014ARA&A..52..251C,2016AJ....151....2D}, and are an important indicator of the scale and intensity of ARs. There is a linear relationship between AR parameters and sunspot parameters and varies across the different cycles and hemispheres. In previous studies, \citet{1980SoPh...66..233B} indicated that the evolution of sunspots and the variation of faculae exhibited high regularity within a given cycle, and this relationship appeared similar in both hemispheres. However, significant differences were observed across different cycles. This contrasted with the differences in the correlation between the two hemispheres found in this study. \citet{1998ApJ...500..958F} showed a clear nonlinear relationship between white-light faculae and sunspot areas, with the variation in their ratio possibly linked to the solar magnetic field generation mechanism. \citet{2012SoPh..277..417H} indicated that there is a significant nonlinear relationship between sunspot number and total solar irradiance (TSI), while its linear relationship with Terrestrial Insolation is stronger than with TSI. Furthermore, \citet{2016SPD....47.1003C,2016SoPh..291.1957C} investigated the relation between the filling factor of magnetic elements, the magnetic element coverage area and sunspot numbers, respectively. They found that daily data fit a quadratic function best, while six-month averages are better described by a linear function. Sunspot activity is closely related to the magnetic field distribution. The AR parameters offer additional insights into changes in sunspot activity, contributing to more accurate predictions of the intensity and cyclical fluctuations of solar activity \citep{1980SoPh...66..233B}. They also reflect deeper complexities and variations in the magnetic field strength \citep{1992ASIC..375....3T,2003A&ARv..11..153S}. Sunspots are not merely accompanying phenomena of ARs, they share underlying physical connection and may be jointly governed by magnetohydrodynamic processes inside the Sun. Studying the relationship between AR parameters and sunspot parameters provides valuable insights for understanding the solar internal magnetic activity and dynamic processes.

The three parameters of AR exhibit hemispheric dominance differences before and after the maximum time of solar cycles, which may be influenced by the reversal of magnetic field polarity. We observe a distinct double-peak structure during SC24 (see Figure \ref{fig:ns}), which is formed by ARs from different hemispheres at different times. \citet{2013ApJ...763...23S} suggested that the occurrence of two or more solar activity peaks in different hemispheres associated with the corresponding differences in polar field reversal times is a common characteristic of solar cycles. The asymmetric polar field reversal is merely a result of the asymmetry in solar activity, indicating that the overall magnetic field evolution of ARs is governed by the differing activity levels in the two hemispheres. We find that the hemispheric ARs exhibit distinct variation patterns across the different solar cycles and parameters (see Table \ref{tab:all}). The similarity between the AR area and flux in the dominant hemisphere during different solar cycles further demonstrates the strong correlation between them. Statistical tests indicate that the N-S asymmetry of the AR parameters is a real observational phenomenon. The current SC25 is underway and existing data show that the southern hemisphere has a slight dominance. The activity trend resembles that of SC23, showing slightly higher intensity than SC24. This is consistent with the predictions of \citep{2022AdSpR..69..798P,2023PASJ...75..691L,2024MNRAS.527.5675S}. As the cycle progresses, this trend of asymmetry is expected to develop further.

The normalized asymmetry index is a useful parameter, provided that the monthly number of ARs is sufficiently large \citep{2022MNRAS.514.1140Z}. Research on the asymmetry index shows that ARs occurrence is uneven in both hemispheres, reflecting the intrinsic complexity of the solar magnetic field (see Figures \ref{fig:number} to \ref{fig:flux}). Moreover, the change in the fitting slope sign before 2013 and after 2014 may be related to the magnetic field polarity reversal near the maximum of SC24 \citep{1959ApJ...130..364B,2015ApJ...798..114S,2019SoPh..294..137P}. This indicates that the generation and reversal mechanisms of the solar magnetic field are not completely symmetric, influenced by the enhancement of local magnetic fields in each hemisphere, which may stem from the nonuniformity of the solar dynamo mechanism. \citet{2007SoPh..245...19J} and \citet{2009RAA.....9..115G} noted that hemispheric asymmetry in the next cycle could be caused by random effects on the dynamo process at the end of the previous cycle. The hemispheric asymmetry of solar ARs is also closely related to different phases of the solar cycle \citep{2006JApA...27..151J,2010SoPh..261..193N}. Furthermore, \citet{2017SCPMA..60a9601C} discussed the extrapolation of the solar dynamo to solar-like stars and the connection between solar cycles and stellar activity cycles. Therefore, the hemispheric asymmetry of solar ARs discussed in this study can also be used to understand the activity behaviors of other stars.

Analysis of cumulative effects reveals that the long-term cumulative trends of the three parameters of AR in different hemispheres exhibit significant differences across the different solar cycles (see Figures \ref{fig:cul} and \ref{fig:abs}). \citet{2023MNRAS.520.3923Z} found that the cumulative trends of different types of CMEs exhibited different patterns during SC23 and SC24. The hemispheric asymmetry of solar activity is not merely a localized phenomenon but exhibits significant long-term and periodic characteristics \citep{2019SoPh..294...64J,2021Ap&SS.366...16J}. 

This study provides a preliminary insight into the patterns of N-S asymmetry in AR parameters. Future observational data will offer more empirical evidence for analyzing hemispheric differences in solar activity and uncovering additional evolutionary trends of ARs. These findings will not only provide a new theoretical framework for solar physics research but also establish a robust scientific foundation for predicting future solar activity and its impact on space weather.

\begin{acknowledgments}
We thank the reviewers for the valuable comments and suggestions on improving the paper. Meanwhile, we thank Ruihui Wang, Jie Jiang, and YuKun Luo for creating the live homogeneous database of solar active regions. This work is supported by the National Nature Science Foundation of China (12463009), the Yunnan Fundamental Research Projects (grant Nos. 202301AV070007, 202401AU070026, 202501AU070154, 202501AT070366), the “Yunnan Revitalization Talent Support Program” Innovation Team Project (grant No. 202405AS350012), the Scientific Research Foundation Project of Yunnan Education Department (2023J0624, 2024Y469, 2025Y0720, 2025Y0721, 2025J0502), and the GHfund A (202407016295).
\end{acknowledgments}

\section*{ORCID iDs}
\noindent 
Yuxia Liu \href{https://orcid.org/0009-0009-5713-5380}{https://orcid.org/0009-0009-5713-5380} \\
Tingting Xu \href{https://orcid.org/0000-0002-9997-9524}{https://orcid.org/0000-0002-9997-9524} \\
Miao Wan \href{https://orcid.org/0000-0001-8865-1497}{https://orcid.org/0000-0001-8865-1497} \\
Linhua Deng \href{https://orcid.org/0000-0003-4407-8320}{https://orcid.org/0000-0003-4407-8320} \\
Xinhua Zhao \href{https://orcid.org/0000-0002-9977-2646}{https://orcid.org/0000-0002-9977-2646} \\
Shiyang Qi \href{https://orcid.org/0000-0002-6957-8009}{https://orcid.org/0000-0002-6957-8009} \\
Nanbin Xiang \href{https://orcid.org/0000-0001-9062-7453}{https://orcid.org/0000-0001-9062-7453}\\
Weihong Zhou \href{https://orcid.org/0009-0003-2287-6441}{https://orcid.org/0009-0003-2287-6441}\\

\bibliography{sample7}{}
\bibliographystyle{aasjournalv7}
\end{document}